\documentclass[prl,twocolumn,preprintnumbers,amsmath,amssymb,a4paper,superscriptaddress,twoside]{revtex4}
\usepackage[version=3]{mhchem} 

\usepackage{graphicx}

\begin{document}

\sloppy
\author{Daniela Kerl\'e}
\email{dkerle@uni-bremen.de}
\affiliation{FB 4 Produktionstechnik, Universit\"at Bremen, Badgasteiner Str.\ 1,  D-28359 Bremen, Germany}
\author{Majid Namayandeh Jorabchi}
\affiliation{Institut f\"ur Chemie, Physikalische und Theoretische Chemie,
  Universit\"at Rostock, Albert-Einstein-Str.\ 21, D-18059 Rostock, Germany}
\author{Ralf Ludwig}
\affiliation{Institut f\"ur Chemie, Physikalische und Theoretische Chemie,
  Universit\"at Rostock, Dr.-Lorenz-Weg~1, D-18059 Rostock, Germany}
\affiliation{Leibniz-Institut f\"ur Katalyse an der
  Universit\"at Rostock, Albert-Einstein-Str.\ 29a, D-18059 Rostock, Germany}
\author{Sebastian Wohlrab}
\affiliation{Leibniz-Institut f\"ur Katalyse an der
  Universit\"at Rostock, Albert-Einstein-Str.\ 29a, D-18059 Rostock, Germany}
\author{Dietmar Paschek}
\email{dietmar.paschek@uni-rostock.de}
\affiliation{Institut f\"ur Chemie, Physikalische und Theoretische Chemie,
  Universit\"at Rostock, Albert-Einstein-Str.\ 21, D-18059 Rostock, Germany}

\title{A Simple Guiding Principle for the  Temperature Dependence of the Solubility
 of Light Gases in Imidazolium-based Ionic Liquids Derived from Molecular Simulations}

 \date{\today}

\begin{abstract}
We have determined the temperature dependence of the solvation behavior of a large collection 
of important light gases in imidazolium-based Ionic Liquids (ILs) with the help of extensive molecular dynamics simulations. The motivation of our study is
to unravel common features of the temperature dependent solvation 
under well controlled conditions, and to provide 
a guidance for cases, where experimental data from different sources
 disagree significantly.
The solubility of molecular hydrogen, oxygen, nitrogen, methane, krypton, 
argon, neon and carbon dioxide in the imidazolium based ILs of type
1-n-alkyl-3-methylimidazolium bis(trifluoromethylsulfonyl)imide
([C$_n$mim][NTf$_2$]) with varying chain lengths $n\!=\!2,4,6,8$ are computed 
for a temperature range between
$300\,\mbox{K}$ and $500\,\mbox{K}$ at $1\,\mbox{bar}$.
By applying Widom's particle insertion technique and Bennet's overlapping
distribution method, we are able to
determine the temperature dependent 
solvation free energies for those selected light gases in
simulated imidazolium based ILs with high statistical accuracy. 
Our simulations demonstrate that the magnitude of the solvation free energy 
of a gas molecule at a chosen reference temperature and
its temperature-derivatives are intimately related with respect to
oneanother. 
We conclude that this ``universal'' 
behavior is rooted in a solvation entropy-enthalpy
compensation effect, which seems to be 
a defining feature of the solvation 
of small molecules in Ionic Liquids. We argue that this feature
is based on a hypothesized funnel-like shape of the free energy landscape of a solvated
gas molecule.
The observations lead to simple analytical relations, determining the
temperature dependence of the solubility data based on the absolute
solubility at a certain reference temperature, which we call
``solvation funnel'' model.
By comparing our results with available experimental data from many sources,
we can show that the ``solvation funnel'' model
 is particularly helpful for providing reliable estimates for  the solvation
behavior of very light gases, such as hydrogen, where conflicting
experimental data exist.
\end{abstract}

\maketitle

\section{Introduction}

Salts with melting points below $100^{\circ}$C
are commonly referred to as Ionic liquids (ILs). These liquids
have several unique properties
\cite{WasserscheidWelton,Angell:2012,Endres:2006,Rogers:2003},
 and are discussed for a wide range of potential applications
 \cite{Smiglak:2007,Plechkova:2008}. 
For the application of ILs in gas separation processes
(e.g. for
flue gas decontamination) it is important to
have access to accurate solubility data.
\cite{Blanchard:1999,Tempel:2008,Huang:2009,Ramdin:2012}.
Even more so, the recently introduced Supported Ionic Liquids Membranes
\cite{Bara:2010,Raeissi:2009,Myers:2008,Ilconich:2007} 
are a promising new tool for separating various mixtures of gases. Of particular
importance, of course, is the ability to separate of H$_2$ and CO$_2$
from gas-streams.

Since now, a wealth of experimental measurements of the infinite dilution properties
for a large number of gases in various ILs have been reported \cite{Lei:2013}.
However, the experimental determination of 
solubilities is particularly difficult for gases with a low molecular weight~\cite{WasserscheidWelton}.
As result, for example, the reported solubility data of hydrogen in ILs~\cite{Finotello:2008,
Jacquemin:2007,Gomes:2007,Kumelan1:2006,Dyson:2003,Jacquemin:2006,Jacquemin:2008,Jacquemin2:2006,Kumelan3:2006,Kumelan:2010,Raeissi:2011} 
are highly inconsistent. 

In addition, also theoretical methods to
determine solubilities of gases in imidazolium-based ILs have been reported in literature:
Data were determined from COSMO based methods~\cite{Shimoyama:2010,Manan:2009,Palomar:2011,Sumon:2011},
equations of state
approaches~\cite{Abildskov:2009,Andreu:2008,Zhang:2008,Andreu:2007},
group theory~\cite{Kim:2007}, 
quantitative structure-property relationship (QSPR) and 
neutral network models~\cite{Oliferenko:2011}, as well as, molecular dynamics (MD) 
and Monte Carlo (MC) simulations \cite{Maginn:2009,Shi1:2008,Shi2:2008,Koeddermann:2008a,Lynden-Bell:2002,Hanke:2003,Deschamps:2004,Shah:2004,Shah:2005,Cadena:2004,Kumelan:2005,Urukova:2005,Kumelan:2007,Deschamps:2005,Wu:2005,Shi:2010,Ghobadi:2011}.
We would like to point out that 
MD and MC simulations have the advantage that they offer the possibility
of both, 
a (semi-)quantitativ
prediction of the solubility,
as well as gaining a fundamental understanding
 of the molecular mechanism for the solvation process.

For the simulation of  imidazolium-based ILs there are different
atomic-detailed molecular force fields available \cite{Dommert:2012}. 
Many of those force-fields are capable of reproducing 
essentially ``static'' properties, such as thermodynamic properties
and structural features quite well. However, most of them
are lacking the ability of 
describing transport properties, such as diffusion coefficients and viscosities
satisfactorily. In this study we used the non-polarizable all-atom
forcefield originally introduced by Lopes \cite{Lopes:2004}, using the
refined parameters 
of K\"oddermann et\,al.~\cite{Koeddermann:2007} to simulate the 
imidazolium-based ILs of type ${\rm [C}_n{\rm mim][NTf_2]}$ (see Figure 1). 
We have shown earlier that a wealth of both, thermodynamical and dynamical properties
of the pure IL could be described
in excellent agreement with experimental data
\cite{Koeddermann:2007,Koeddermann:2008a} by
using this model.
In addition, this modified forcefield was also capable describing the solvation behavior of 
noble gases \cite{Paschek:2008} and carbon dioxide \cite{Kerle:2009}
very satisfactorily.
By accurately determining temperature dependent solvation properties, we
 could demonstrate that the entropy contribution to the solvation
free energy plays an important role in the solvation process, not unlike the
hydrophobic hydration of small apolar particles in liquid
water \cite{PrattRev:2002,Southall:2002,Widom:2003,Chandler:2005}.
Moreover, also an
entropy-driven ``solvophobic interaction'' of apolar particles could
be observed in ILs \cite{Paschek:2008}, indicating
that specific solvent-mediated interactions could play an important
role in ILs.

Here we focus on the infinite dilution properties of ``light gases'', 
like carbon dioxide, oxygen, nitrogen, methane, argon, neon, and
hydrogen in imidazolium based Ionic Liquids. The chosen gases cover a spectrum from
very weakly interacting gases, such as hydrogen, to moderately strong
interacting molecules, such as carbon dioxide.
We apply Widom's particle insertion  technique for calculating
temperature dependent solvation free energies and solubilities
of these gases in imidazolium-based ionic liquids of the type
${\rm[C}_n{\rm mim][NTf_2]}$ with
varying chain lengths.
$n\!=\!2,4,6,8$. In addition, to validate these calculations we also use 
Bennett's overlapping distribution method for selected examples.
The calculated Henry constants are
compared with available experimental data. The temperature behavior of the
solubility as well as its dependence of the alkyl-chain lengths in the
imidazolium cations is determined and discussed.
The motivation of our study is
to reveal common ``universal'' features of the temperature dependent solvation 
under well controlled conditions, and to provide 
a reasonable guidance for cases, where experimental data from different sources
 disagree significantly.
\begin{figure}[!t]
  \centering
  \footnotesize
  \centering
  \includegraphics[angle=0,width=7.0cm]{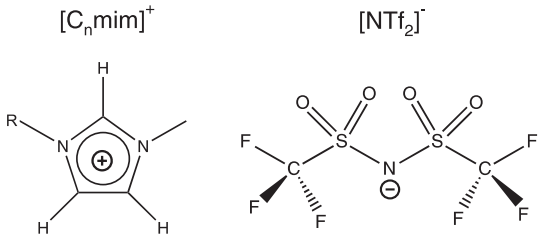}
  \caption{Schematic representation of the studied ionic liquids
  1-n-alkyl-3-methylimidazolium bis(trifluoromethyl\-sulfonyl)imide
  ${\rm[C}_n{\rm mim][NTf_2]}$.}
\end{figure}

\section{Experimental Section}

\subsection{Molecular Dynamics Simulations}

We perform constant pressure (NPT) MD simulations of imidazolium based ILs of
the type  [C$_n$mim][NTf$_2$] for different chain lengths $n=2,4,6,8$
at a pressure of $1\,\mbox{bar}$, covering
a broad temperature range between  $300\,\mbox{K}$ and $500\,\mbox{K}$.
All simulated systems are composed of $343$ ion pairs, applying the
forcefield of Lopes et al. \cite{Lopes:2004} with refined potential parameters
according to
 K\"oddermann et\,al.~\cite{Koeddermann:2007}.
An additional minor modification from the simulation setup used in previous studies
\cite{Koeddermann:2007,Koeddermann:2008a} is that {\em all}
bond-lengths were kept fixed.
A cubic simulation box was used, and the system size with 343 ion
pairs was chosen large enough 
that thermodynamical properties do not depend on the system-size, as it was
reported by Wittich et\,al.~\cite{Wittich:2010}
for a system of 125 ion pairs of [C$_4$mim][PF$_6$].
For the solutes, various models were employed.
For hydrogen, the potential of Potkowski et\,al.~\cite{Patkowski:2008} was used.
Potential modes reported by Guillot et\,al.~\cite{Guillot:93} were
employed to describe the noble gases and methane.
Nitrogen was described by the Potential of Potoff et\,al.~\cite{Potoff:2001} and oxygen 
by the Potential of Hansen et\,al.~\cite{Hansen:2007}.
Finally, carbon dioxide the
EPM2-model of Harris and Yung~\cite{Harris:1995} was used in a modified way as described before \cite{Kerle:2009}. All parameters describing the solutes are give in
Table \ref{tab:paras}.
\begin{table}[!t]
  \centering
  \renewcommand{\arraystretch}{1.0}
  \renewcommand{\tabcolsep}{0.19cm}
  \caption{Forcefield parameters describing the studies gaseous solutes.
           Given are the Lennard-Jones parameter for the solute-solute site-site               
           interactions $\sigma_{ii}$, and $\varepsilon_{ii}$, the partial charges
           $q_{i}$, as well as the intramolecular bond-lengths $d$. Lennard-Jones 
           cross parameters
           for the solute-solvent interactions were obtained from
           Lorentz-Berthelot combination rules.}
  \label{tab:paras}
  \begin{tabular}{lccccc} \hline\hline \\[-6pt]
 ~    & ~   & $\varepsilon_{ii}\cdot {\rm k}_B^{-1}/\mbox{K}$ & $\sigma_{ii}/\mbox{\AA}$  & $q_i/|e|$ & $d/ \mbox{\AA}$ \\[6pt] \hline \\[-6pt] 
H$_2$ \cite{Patkowski:2008} & ~   & $35.45$ & $3.46$  & ~        & ~ \\
Ne \cite{Guillot:93}   & ~   & $18.6$  & $3.035$ & ~        & ~ \\
Ar \cite{Guillot:93} & ~   & $125.0$ & $3.415$ & ~        & ~ \\
Kr \cite{Guillot:93} & ~ & $169.0$ & $3.675$ &  ~ & ~ \\
CH$_4$ \cite{Guillot:93} & ~  & $147.4$ & $3.73$  & ~        & ~ \\
N$_2$ \cite{Potoff:2001} & N   & $36.0$  & $3.31$  & $-0.482$ & $1.10$ \\
~     & COM & ~       & ~       & $+0.964$ & ~ \\
O$_2$ \cite{Hansen:2007} & O   & $49.048$ & $3.013$ & $-0.123$ & $1.21$ \\
~     & COM & ~       & ~       & $+0.246$ & ~ \\
CO$_2$ \cite{Harris:1995} & C  & $28.129$ & $2.757$ & $+0.6512$ & $1.149$\\
~     & O   & $80.507$ & $3.033$ & $-0.3256$ & ~ \\[6pt]\hline\hline\\[-6pt]
  \end{tabular}
\end{table}

All simulations reported here were performed with the Gromacs simulation
program \cite{gmxpaper}. The preparation of topology files, as well as the
data analysis were performed with the most recent version of the
MOSCITO suite of programs \cite{MOSCITO}. 
Production runs of $10\,\mbox{ns}$ length were employed for every temperature, 
starting from previously well equilibrated configurations. 
The Nos\'e-Hoover thermostat~\cite{Nose:1984,Hoover:1985}
and the Parrinello-Rahman barostat~\cite{Parrinello:1981,Nose:1983} with
coupling times $\tau_T\!=\!1.0\,\mbox{ps}$ and $\tau_p\!=\!2.0\,\mbox{ps}$
were used to control constant temperature and pressure ($1\,\mbox{bar}$) conditions.
The electrostatic interactions were treated by particle mesh Ewald summation \cite{Essmann:1995}.
A real space cutoff of $1.2\,\mbox{nm}$ was employed, 
and a mesh spacing of approximately $0.12\,\mbox{nm}$ (4th order interpolation)
had been used to determine the reciprocal lattice contribution.
The Ewald convergence parameter was set to a relative accuracy of the Ewald sum of $10^{-5}$.
Lennard-Jones cutoff corrections for energy and pressure were considered. 
A $2\,\mbox{fs}$ timestep was used in all simulations, 
and every 25th steps a configuration was saved.
Distance constraints were solved by means of the SHAKE procedure \cite{Ryckaert:1977}.
The thermodynamic properties of the simulated ionic liquids are essentially
identical to the properties reported in Ref. \cite{Kerle:2009}.

\subsection{Infinite Dilution Properties}

The solubility of a solute A in a solvent B is 
conventiently described by the Ostwald
coefficient $L^{l/g}\!=\!\rho^l_{\rm A}/\rho^g_{\rm A}$, where
$\rho^l_{\rm A}$ and
$\rho^g_{\rm A}$ are the number densities of the solute in the
liquid and the gas phase of component B, respectively, when both
phases are in equilibrium. Alternatively, the solubility of
solute A can be
expressed in terms of the inverse Henry's constant $k_{\rm H}^{-1}$.
The relationship between  Henry's constant and the 
excess chemical potential $\mu_{\rm ex,A}^l$ in the liquid phase
is given by~\cite{Kennan:90}
\begin{equation}
k_{\rm H}^{-1}\!=\! \exp\left[-\beta\,\mu_{\rm ex,A}^l\right]/
    \left( \rho^l_{\rm IL} {\rm R} T\right) \,,
\label{eq:solubility}
\end{equation}
where $\beta\!=\!1/{\rm k_B} T$ and $\rho^l_{\rm IL}$ represents the number density of
ion pairs in the IL solvent. 

According to Widom's potential distribution theorem~\cite{Widom:63,Beckbook},
the excess chemical potential $\mu_{\rm ex}$ can be computed as volume
 weighted ensemble average
\begin{equation}
\mu_{\rm ex}\;= \; - {\rm k_B}T \ln \left< V \exp(-\beta\,\Phi)
\right>/\left<V\right>\;.
\end{equation}
Here $V$ is the volume of the simulation box,
and $\Phi$ is the energy of a gas molecule inserted at a
random position with a random orientation. 
The brackets $\left< \ldots \right>$ indicate isothermal-isobaric
averaging over many configurations, as well as averaging over many 
insertions.

As control, we also determine the excess chemical potential from energy
histograms~\cite{Bennett:1976,Shing:1983}
computed for the energy change $\Delta U=U(N+1)-U(N)$ associated with the insertion 
$p_0(\Delta U)$
and removal $p_1(\Delta U)$ of a $(N+1)$th gas molecule from the
constant pressure (NPT) simulation. 
The two distribution functions are related according to
\begin{eqnarray}
p_1(\Delta U)
& = &
\frac{Q(N,P,T)}{Q(N+1,P,T)}\;
\frac{\left<V\right>}{\Lambda^{3}} \;\times \nonumber\\
&  &
\exp(-\beta \Delta U)\;
p_0(\Delta U) \;,
\end{eqnarray}
using the definition of the {\em ideal} and {\em excess} part
of the chemical potential $\mu$ referring to
the ideal gas state with the same  average number density~\cite{FrenkelSmit},
 a relation between the two distribution functions and the excess chemical
potential is obtained, which is analogous to the 
expression for the canonical ensemble~\cite{FrenkelSmit}
\begin{eqnarray}
\ln p_1(\Delta U) - \ln p_0(\Delta U)
& = &
\beta \mu_{\rm ex} -\beta \Delta U \;.
\end{eqnarray}
The only difference is the necessity of {\em volume-weighting}
in the calculation of the $p_0(\Delta U)$-distribution function~\cite{Paschek:2004:1}.
For reasons of convenience we define functions $f_0$ and $f_1$ according to
\begin{eqnarray}
f_0(\Delta U)
& = & \beta^{-1} \ln p_0(\Delta U) - \frac{\Delta U}{2} \;\;\;,\hspace*{2em}\mbox{and}\nonumber\\
f_1(\Delta U)
& = & \beta^{-1} \ln p_1(\Delta U) + \frac{\Delta U}{2} \;\;\;,\nonumber
\end{eqnarray}
such that
\begin{eqnarray}
\mu_{\rm ex} & = &
f_1(\Delta U)-f_0(\Delta U)\;.
\end{eqnarray}

All computed energies are based on the minimum image and include a reaction
field correction similar to Roberts and Schnitker \cite{Roberts:94}. Cut-off
corrections for the dispersion interactions are included \cite{Paschek:2004:2}.

A total of $2\times 10^5$ configurations were analysed for each IL and for every temperature.
Each configuration was sampled by $10^3$ random insertions to determine the 
$f_0$-functions. The energies computed for those insertions has also 
been used to determine ``Widom-estimates'' for the excess chemical
potentials.
We would like to point out that the values computed
from particle insertions are found to lie within the statistical
uncertainty of the data from the overlapping distribution theory.
 Note that the choice of the
sampling rate is a critical parameter for successfully computing
the chemical potentials via Widom's insertion technique.
By reducing the sampling rate significantly, we denote a systematic deviation
of the ``Widom-estimate'' from the data obtained via the overlapping distribution
method. This effect was observed by us for sampling rates 
being about two orders of magnitude lower than the rates reported here.
All ``converged'' computed Henry coefficients are shown in Table \ref{tab:henry}.

From the 
temperature dependence of the computed solvation free energy for infinite
dilution, we can comment on the behavior 
of the first and second derivatives of free energy with
respect to temperature.
So the solvation entropies, enthalpies, and heat capacities
are obtained from fits of
the data to a second order expansion of the solvation
free energy around reference state
($T^\circ\!=\!298\,\mbox{K}$ at $P^\circ\!=\!1\,\mbox{bar}$)
according to
\begin{eqnarray}
\label{eq:cpfit}
\mu_{\rm ex}(T)&=&
\mu^\circ_{\rm ex} - s^\circ_{\rm ex} (T-T^\circ) \\\nonumber
&& -c_{P,\rm ex}\left[ T(\ln T/T^\circ -1) + T^\circ \right]\;.
\end{eqnarray}
Here $\mu^\circ_{\rm ex}$ and $s^\circ_{\rm ex}$
represent the solvation free energy and solvation
entropy at the reference state, respectively.
According to the second order expansion, the solvation 
heat capacity $c_{P,\rm ex}$ is assumed to be constant over the considered temperature range.
The fitted parameters are provided in Table \ref{tab:chempot_fit}.


\section{Results and Discussion}

\begin{table*}[!ht]
\centering
\renewcommand{\tabcolsep}{0.9cm}
\caption{Calculated Henry constants $k_{\rm H}$ for various gaseous components
  dissolved in
  in imidazolium based ionic liquids of type ${\rm [C}_n{\rm mim][NTf_2]}$. 
  All data were obtained from MD simulations at $1\,\mbox{bar}$  and describe 
  the  infinite dilution limit according to
  $k_{\rm H}\!=\! \exp\left[ \beta\,\mu_{\rm ex,\rm Gas}^l\right]\times\rho^l_{\rm IL} R
  T $ \cite{Kennan:90}. The ion-pair densities are computed from fitted second order polynomial $\rho^l_{\rm IL}$ \cite{densityfit}.}
\label{tab:henry}
  \begin{tabular}{lc|r@{$\pm$}l r@{$\pm$}l r@{$\pm$}l r@{$\pm$}l} \\ \hline\hline 
~ & $T/\mbox{K}$ & \multicolumn{8}{c}{$k_{\rm H} / \mbox{bar}$} \\  \hline
 ~   & ~  & \multicolumn{2}{c}{[$\mbox{C}_2$mim]} & \multicolumn{2}{c}{[$\mbox{C}_4$mim]}  & \multicolumn{2}{c}{[$\mbox{C}_6$mim]} & \multicolumn{2}{c}{[$\mbox{C}_8$mim]}  \\ \hline 
$\mbox{CO}_2$ & 300  &  $34$&$3$    &  $28$&$3$   &  $29$&$3$   &  $22$&$3$   \\
~ & 350  &  $87$&$6$    &  $73$&$5$   &  $63$&$4$   &  $56$&$6$   \\
~ & 400  &  $140$&$8$   &  $121$&$6$  &  $107$&$6$  &  $98$&$9$   \\
~ & 450  &  $207$&$10$  &  $171$&$9$  &  $152$&$8$  &  $139$&$11$ \\
~ & 500  &  $262$&$12$  &  $224$&$11$ &  $191$&$9$  &  $185$&$13$ \\\hline

$\mbox{Kr}$& 300  &  $188$&$5$    &  $143$&$6$   &  $119$&$6$   &  $89$&$6$   \\
~& 350  &  $266$&$4$  &  $210$&$3$ &  $175$&$2$  &  $139$&$3$   \\
~& 400  &  $327$&$3$  &  $261$&$3$ &  $221$&$2$  &  $182$&$3$   \\
~& 450  &  $377$&$2$  &  $306$&$2$ &  $257$&$1$  &  $214$&$1$ \\
~& 500  &  $411$&$2$  &  $337$&$3$ &  $284$&$2$  &  $240$&$1$ \\\hline

$\mbox{CH}_4$ & 300  &  $300$&$10$  &  $224$&$15$  &  $214$&$15$   &  $142$&$10$   \\
~& 350  &  $381$&$7$   &  $300$&$7$   &  $277$&$6$    &  $203$&$5$    \\
~& 400  &  $441$&$3$   &  $351$&$4$   &  $321$&$4$    &  $249$&$3$    \\
~& 450  &  $486$&$5$   &  $391$&$3$   &  $353$&$3$    &  $278$&$2$    \\
~& 500  &  $509$&$4$   &  $415$&$3$   &  $373$&$3$    &  $299$&$2$    \\\hline

$\mbox{Ar}$ & 300  &  $436$&$9$   &  $350$&$12$  &  $318$&$14$   &  $242$&$15$   \\
~& 350  &  $508$&$6$   &  $418$&$6$   &  $378$&$4$    &  $303$&$6$    \\
~& 400  &  $548$&$4$   &  $455$&$5$   &  $410$&$4$    &  $340$&$4$    \\
~& 450  &  $571$&$3$   &  $478$&$3$   &  $427$&$2$    &  $356$&$2$    \\
~& 500  &  $577$&$2$   &  $486$&$3$   &  $431$&$3$    &  $365$&$2$    \\\hline

$\mbox{O}_2$ & 300  &  $652$&$22$  &  $523$&$21$  &  $456$&$23$   &  $378$&$26$   \\
~& 350  &  $720$&$7$   &  $596$&$8$   &  $517$&$5$    &  $447$&$10$    \\
~& 400  &  $744$&$5$   &  $619$&$7$   &  $540$&$5$    &  $480$&$6$    \\
~& 450  &  $750$&$5$   &  $629$&$4$   &  $545$&$3$    &  $482$&$3$    \\
~& 500  &  $732$&$4$   &  $621$&$5$   &  $536$&$3$    &  $477$&$3$    \\\hline

$\mbox{N}_2$ & 300  &  $1048$&$47$  &  $858$&$42$   &  $766$&$46$   &  $651$&$53$   \\
~& 350  &  $1062$&$15$  &  $898$&$14$   &  $795$&$12$   &  $707$&$20$    \\
~& 400  &  $1036$&$7$   &  $875$&$11$   &  $775$&$9$    &  $703$&$11$    \\
~& 450  &  $991$&$8$    &  $844$&$7$    &  $741$&$4$    &  $666$&$6$    \\
~& 500  &  $942$&$4$    &  $802$&$7$    &  $699$&$4$    &  $629$&$5$    \\\hline

$\mbox{Ne}$ & 300  &  $3418$&$63$  &  $3051$&$88$   &  $2898$&$110$   &  $2752$&$172$   \\
~& 350  &  $2507$&$23$   &  $2222$&$25$   &  $2088$&$16$   &  $1964$&$36$    \\
~& 400  &  $1943$&$13$   &  $1716$&$10$   &  $1587$&$13$   &  $1480$&$8$    \\
~& 450  &  $1570$&$12$   &  $1388$&$8$    &  $1267$&$5$    &  $1185$&$10$    \\
~& 500  &  $1318$&$6$    &  $1147$&$5$    &  $1047$&$5$    &  $977$&$7$    \\\hline

$\mbox{H}_2$ & 300  &  $3875$&$98$  &  $3286$&$121$   &  $3313$&$157$   &  $2718$&$173$   \\
~& 350  &  $2867$&$42$   &  $2466$&$44$   &  $2415$&$33$   &  $2066$&$42$    \\
~& 400  &  $2216$&$16$   &  $1906$&$23$   &  $1835$&$19$   &  $1610$&$21$    \\
~& 450  &  $1805$&$22$   &  $1552$&$10$   &  $1462$&$7$    &  $1279$&$9$    \\
~& 500  &  $1510$&$5$    &  $1296$&$10$   &  $1190$&$13$   &  $1058$&$8$    \\\hline\hline

  \end{tabular}
\end{table*}

\begin{table*}[!t]
\renewcommand{\arraystretch}{1.0}
\renewcommand{\tabcolsep}{0.8cm}
\caption{Thermodynamic parameters describing the temperature dependence
  of the solvation free energies $\mu_{\rm ex}(T)$ of the indicated solutes according to 
  a second order expansion around a thermodynamic reference state, following Eq.\ \ref{eq:cpfit}. The chosen reference state has been set to $T^\circ\!=\!298\,\mbox{K}$ at $P^\circ\!=\!1\,\mbox{bar}$.}
\label{tab:chempot_fit}
  \begin{tabular}{lc|cccc} \\\hline\hline
 ~ &  ~   & [$\mbox{C}_2$mim] & [$\mbox{C}_4$mim] & [$\mbox{C}_6$mim] & [$\mbox{C}_8$mim] \\\hline
$\mbox{CO}_2$ & $\mu^\circ_{\rm ex}/\mbox{kJ}\,\mbox{mol}^{-1}$          & $-2.57$ & $-2.79$ & $-2.45$ &  $-2.89$  \\
~ & $s^\circ_{\rm ex}/\mbox{J}\mbox{K}^{-1}\,\mbox{mol}^{-1}$& $-39.1$ & $-40.1$ & $-33.8$ &  $-39.2$ \\
~ & $h^\circ_{\rm ex}/\mbox{kJ}\,\mbox{mol}^{-1}$          &  $-14.2$ & $-14.7$ & $-12.5$ & $-14.6$   \\
~ & $c_{P,\rm ex}/\mbox{J}\mbox{K}^{-1}\,\mbox{mol}^{-1}$ & $45$ & $48$ & $34$ & $43$  \\\hline 

$\mbox{Kr}$ & $\mu^\circ_{\rm ex}/\mbox{kJ}\,\mbox{mol}^{-1}$          & $1.66$ & $1.24$ & $1.09$ &  $0.56$  \\
~ & $s^\circ_{\rm ex}/\mbox{J}\mbox{K}^{-1}\,\mbox{mol}^{-1}$& $-20.5$ & $-21.3$ & $-20.9$ &  $-23.1$ \\
~ & $h^\circ_{\rm ex}/\mbox{kJ}\,\mbox{mol}^{-1}$          &  $-4.45$ & $-5.09$ & $-5.13$ & $-6.32$   \\
~ & $c_{P,\rm ex}/\mbox{J}\mbox{K}^{-1}\,\mbox{mol}^{-1}$ & $19.9$ & $21.9$ & $21.7$ & $26.8$  \\\hline 

$\mbox{CH}_4$ & $\mu^\circ_{\rm ex}/\mbox{kJ}\,\mbox{mol}^{-1}$          & $2.78$ & $2.36$ & $2.55$ &  $1.73$  \\
~ & $s^\circ_{\rm ex}/\mbox{J}\mbox{K}^{-1}\,\mbox{mol}^{-1}$& $-18.5$ & $-19.7$ & $-18.4$ &  $-21.8$ \\
~ & $h^\circ_{\rm ex}/\mbox{kJ}\,\mbox{mol}^{-1}$          &  $-2.72$ & $-3.50$ & $-2.92$ & $-4.76$   \\
~ & $c_{P,\rm ex}/\mbox{J}\mbox{K}^{-1}\,\mbox{mol}^{-1}$ & $17.6$ & $20.8$ & $18.6$ & $27.0$  \\\hline 

$\mbox{Ar}$ & $\mu^\circ_{\rm ex}/\mbox{kJ}\,\mbox{mol}^{-1}$          & $3.76$ & $3.48$ & $3.54$ &  $3.06$  \\
~ & $s^\circ_{\rm ex}/\mbox{J}\mbox{K}^{-1}\,\mbox{mol}^{-1}$& $-16.2$ & $-16.7$ & $-16.6$ &  $-18.4$ \\
~ & $h^\circ_{\rm ex}/\mbox{kJ}\,\mbox{mol}^{-1}$          &  $-1.07$ & $-1.51$ & $-1.42$ & $-2.41$   \\
~ & $c_{P,\rm ex}/\mbox{J}\mbox{K}^{-1}\,\mbox{mol}^{-1}$ & $16.9$ & $18.4$ & $19.3$ & $23.5$  \\\hline 

$\mbox{O}_2$ & $\mu^\circ_{\rm ex}/\mbox{kJ}\,\mbox{mol}^{-1}$          & $4.76$ & $4.48$ & $4.43$ &  $4.17$  \\
~ & $s^\circ_{\rm ex}/\mbox{J}\mbox{K}^{-1}\,\mbox{mol}^{-1}$& $-16.8$ & $-17.2$ & $-17.1$ &  $-19.0$ \\
~ & $h^\circ_{\rm ex}/\mbox{kJ}\,\mbox{mol}^{-1}$          &  $-0.23$ & $-0.66$ & $-0.67$ & $-1.50$   \\
~ & $c_{P,\rm ex}/\mbox{J}\mbox{K}^{-1}\,\mbox{mol}^{-1}$ & $19.0$ & $20.0$ & $20.7$ & $26.0$  \\\hline

$\mbox{N}_2$ & $\mu^\circ_{\rm ex}/\mbox{kJ}\,\mbox{mol}^{-1}$          & $5.95$ & $5.72$ & $5.73$ &  $5.53$  \\
~ & $s^\circ_{\rm ex}/\mbox{J}\mbox{K}^{-1}\,\mbox{mol}^{-1}$& $-15.5$ & $-16.4$ & $-16.2$ &  $-18.4$ \\
~ & $h^\circ_{\rm ex}/\mbox{kJ}\,\mbox{mol}^{-1}$          &  $1.34$ & $0.84$ & $0.90$ & $0.04$   \\
~ & $c_{P,\rm ex}/\mbox{J}\mbox{K}^{-1}\,\mbox{mol}^{-1}$ & $17.0$ & $20.0$ & $21.0$ & $27.6$  \\\hline 

$\mbox{Ne}$ & $\mu^\circ_{\rm ex}/\mbox{kJ}\,\mbox{mol}^{-1}$          & $8.89$ & $8.90$ & $9.07$ &  $9.13$  \\
~ & $s^\circ_{\rm ex}/\mbox{J}\mbox{K}^{-1}\,\mbox{mol}^{-1}$& $-5.66$ & $-5.91$ & $-5.36$ &  $-4.96$ \\
~ & $h^\circ_{\rm ex}/\mbox{kJ}\,\mbox{mol}^{-1}$          &  $7.20$ & $7.14$ & $7.47$ & $7.65$   \\
~ & $c_{P,\rm ex}/\mbox{J}\mbox{K}^{-1}\,\mbox{mol}^{-1}$ & $10.1$ & $12.3$ & $11.7$ & $10.8$  \\\hline 

$\mbox{H}_2$ & $\mu^\circ_{\rm ex}/\mbox{kJ}\,\mbox{mol}^{-1}$          & $9.19$ & $9.09$ & $9.40$ &  $9.11$  \\
~ & $s^\circ_{\rm ex}/\mbox{J}\mbox{K}^{-1}\,\mbox{mol}^{-1}$& $-7.29$ & $-7.72$ & $-7.29$ &  $-8.99$ \\
~ & $h^\circ_{\rm ex}/\mbox{kJ}\,\mbox{mol}^{-1}$          &  $7.02$ & $6.79$ & $7.23$ & $6.43$   \\
~ & $c_{P,\rm ex}/\mbox{J}\mbox{K}^{-1}\,\mbox{mol}^{-1}$ & $11.3$ & $13.4$ & $14.9$ & $19.0$  \\\hline\hline
  \end{tabular}
\end{table*}

\begin{figure*}
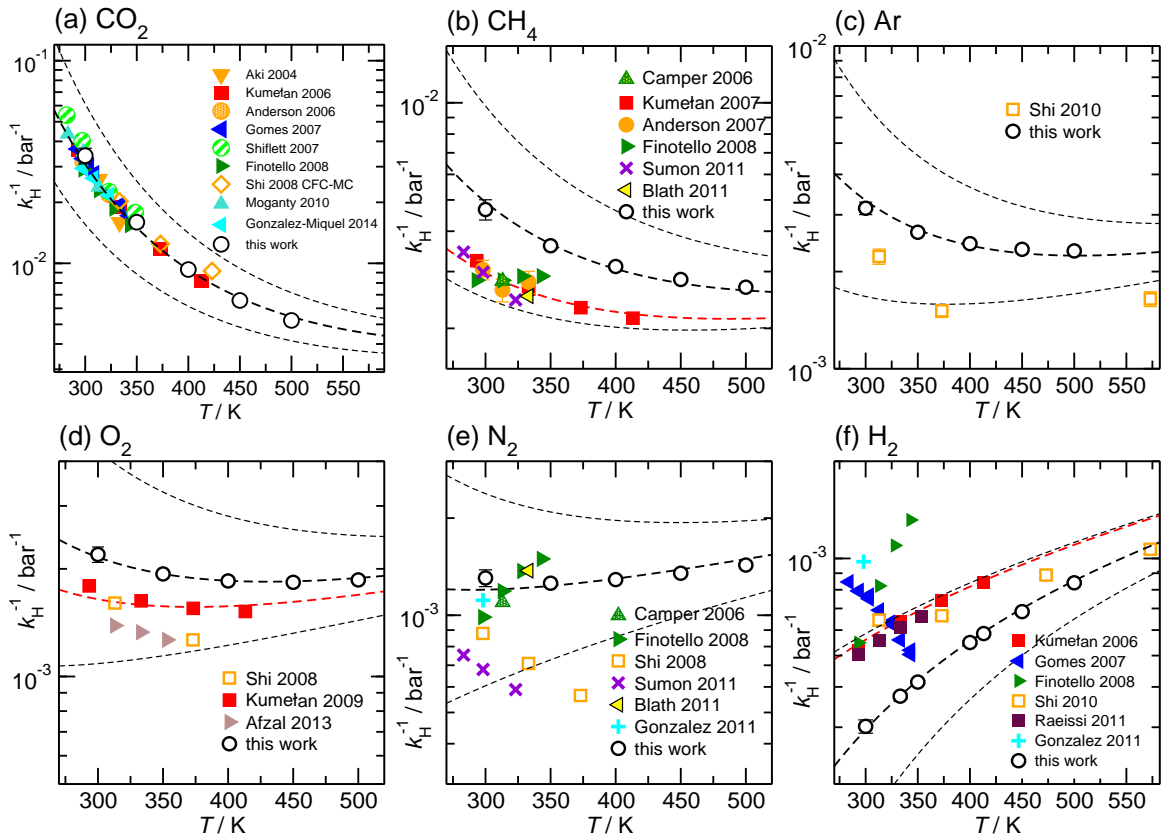

  \centering 
  \footnotesize 
  \centering 
  \includegraphics[angle=0,width=5.0cm]{FIG02a.eps}
  \includegraphics[angle=0,width=5.0cm]{FIG02b.eps}
  \includegraphics[angle=0,width=5.0cm]{FIG02c.eps}

  \includegraphics[angle=0,width=5.0cm]{FIG02d.eps}
  \includegraphics[angle=0,width=5.0cm]{FIG02e.eps} 
  \includegraphics[angle=0,width=5.0cm]{FIG02f.eps}

  \caption{Solubilities $k_{\rm H}^{-1}\!=\! \exp\left[-\beta\,\mu_{\rm ex,\rm Gas}^l\right]/
    \rho^l_{\rm IL} R T $ of selected gases in
    in $\rm[C_6mim][NTf_2]$. Shown are data for  
     a) $\mbox{CO}_2$, 
     b) $\mbox{CH}_4$, 
     c) Ar,
     d) $\mbox{O}_2$, 
     e) $\mbox{N}_2$, and
     f) $\mbox{H}_2$
    at $1\,\mbox{bar}$. 
    The filled symbols represent experimental data given according to the indicated sources.   
    Open symbols specify data from to molecular simulations, while
    crosses represent data obtained from alternative theoretical predictions.
    All dashed lines show theoretical predictions based on our ``solvation funnel'' model.
    The upper and lower black thin dashed lines are
    temperature-predictions for solubilities twice and half of the reference-value
    used for our MD simulation-data (given as black thick dashed line).
    The red dashed line represents a ``solvation funnel prediction'' for the 
    experimental solubility data of 
    Kume{\l}an et al.\
    References for the experimental and theoretical data are given in the text.}	
    \label{fig:solubility}
\end{figure*}

We have computed the solvation free energies $\mu_{\rm ex}$ and
the associated Henry constants $k_{\rm H}$
for
 $\mbox{CO}_2$, $\mbox{O}_2$,  $\mbox{N}_2$, $\mbox{CH}_4$, Kr, Ar, Ne and $\mbox{H}_2$ at infinite
dilution  from MD simulations for the four Ionic Liquids ${\rm [C}_n{\rm mim][NTf_2]}$ with
$n\!=\!2,4,6,8$, employing the sampling techniques  discussed
in the previous section. 
The density data for the simulated ionic liquids, necessary to interconvert
Henry constants $k_{\rm H}$ and solvation free energies $\mu_{\rm ex}$ can be found
in Ref. \cite{Kerle:2009}.
All computed Henry constants are summarised in Table \ref{tab:henry}.
The temperature dependence of the corresponding solvation free 
energies $\mu_{\rm ex}(T)$ has been fitted to a second order expansion
around a reference state
following Eq.\ \ref{eq:cpfit}.
The parameters obtained for a reference temperature of $T^\circ=298\,\mbox{K}$
are given in Table \ref{tab:chempot_fit}.
In Figure \ref{fig:solubility} a section of computed
solubilities (here given as inverse Henry's constants) 
are compared
with available experimental and theoretical data.
For reasons of clarity, we restrict this comparison to data 
based on solvation in $\rm [C_6mim][NTf_2]$. 
Most experimental data sets are only available for 
this particular IL,  since it had been selected as the reference 
compound for an IUPAC experimental validation project \cite{IUPAC1,IUPAC2}.
The remainder of this section is organised as follows: First we will 
compare our simulated solubility data with available experimental and
theoretical data. A short discussion of the available solubility 
data is given for every gas in detail.
The following sections will then focus on a systematic
rationalization of the effect of varying the alkane chain length, and
of changing the temperature by introducing the concept of
the ``solvation funnel''.

\subsection{Comparison with Available Experimental and Theoretical Data}

\subsubsection{Carbon Dioxide}

For carbon dioxide we have shown  previously \cite{Kerle:2009} that all 
available experimental and theoretical data are in excellent agreement with our simulation results.
Particularly, the
 temperature dependence is reflected very well by the simulations \cite{Kumelan:2006,Gomes:2007,Finotello:2008,Moganty:2010}. Moreover, we could recently show that small
differences with respect to a few experimental data sets could be explained
by water content in the samples \cite{Kerle:2013}.
The data calculated for $\mbox{CO}_2$  by Sumon et\,al.\ \cite{Sumon:2011} using COSMO-RS,
and Wu et al.\ \cite{Wu:2005}, as well as Shi et\,al.~\cite{Shi1:2008} using
molecular simulation techniques,
show the same temperature trend for carbon dioxide and are all very close
to the experimental data.

\subsubsection{Methane}

Experimental data for the solubility of methane are available from various sources. 
Kume{\l}an et\,al.~\cite{Kumelan:2007} have examined
methane in $\rm [C_6mim][NTf_2]$ over a large temperature range. In addition, the data of 
Finotello et\,al.~\cite{Finotello:2008} are shown, who 
investigated $\rm [C_2mim][NTf_2]$ and $\rm [C_6mim][NTf_2]$.
Moreover, we also show the data according to
 Camper et\,al.~\cite{Camper2:2006}, as well as
  Anderson et\,al.~\cite{Anderson:2007}, and Blath et\,al.~\cite{Blath:2011} for $\rm [C_6mim][NTf_2]$.
Our data predict a systematically higher solubility compared
with the mostly consistent experimental data sets, 
but are well in 
agreement with the temperature slope of 
the data of Kume{\l}an et\,al.\ 
The temperature dependencies reported by Finotello and Anderson, however,
are clearly not compatible with our findings and the
data of Kume{\l}an et\,al.\ 
We would like to point out, that we did not perform any
adjustment to the force field parameters to improve
the solute-solvent interaction. The Lennard-Jones parameters of Guillot et\,al. apparently
 overestimate the interaction between the solvent and solute.
The data calculated for $\mbox{CH}_4$
 by Sumon et\,al.\ using COSMO-RS \cite{Sumon:2011} 
show the same temperature trend for methane as our simulations.

\subsubsection{Noble Gases}

For the case of krypton there is only one 
experimental dataset available, published by Afzal et\,al.~\cite{Afzal:2013}
in 2013 (not shown here). It shows the same temperature trend as our simulated results.
For the case of argon and neon are, to our knowledge, no experimental data available. 
in 2010 Shi et\,al.~\cite{Shi:2010} published Henry constants 
obtained from computer simulations of argon
in $\rm [C_6mim][NTf_2]$. Their data are in the same decade compared to ours, and 
the temperature dependence is similar, but their computed
solubilities are slightly smaller.

\subsubsection{Oxygen}

For the case of oxygen, there are 
only few experimental solubility data sets available.
This is likely due to the experimental challenges associated with the
use of oxygen.
Our data apparently slightly overestimates the solubility of oxygen, but seems
to agree particularly well with the slope of the temperature dependence 
reported by Kume{\l}an et\,al.~\cite{Kumelan1:2009} in $\rm [C_6mim][NTf_2]$. 
The temperature dependent data of Anthony et\,al.~\cite{Anthony:2005} for $\rm [C_4mim][NTf_2]$ (not shown here), however, suggest a significantly stronger
temperature dependence, in accordance with the recently published values of Afzal et\,al.~\cite{Afzal:2013}.
The simulation-based 
data of Shi et\,al.~\cite{Shi1:2008} seem to be placed right in the middle
between the data of Afzal et\,al.\ and  Kume{\l}an et\,al.\

\subsubsection{Nitrogen}

The solubility of nitrogen has been studied by several groups. 
Unfortunately, the temperature trends of
the available experimental and theoretical data sets seem to be quite inconsistent.
The data of Camper et\,al.~\cite{Camper2:2006} and Finotello et\,al.~\cite{Finotello:2008} in $\rm [C_2mim][NTf_2]$ and $\rm [C_6mim][NTf_2]$,
as well as Blath et\,al.~\cite{Blath:2011} (measured at $60^{\circ}$C in $\rm [C_6mim][NTf_2]$)
are very close with respect to each other. Our data is in the same
range as the experimental data, however, we do not observe a significant increase
of the computed solubilities with increasing temperature.
Both, the data calculated for $\mbox{N}_2$  by Sumon et\,al.\ \cite{Sumon:2011} using
COSMO-RS \cite{Sumon:2011}, and  Shi et\,al.~\cite{Shi1:2008} using
computer simulation techniques
show a significantly different temperature dependence compared
to the experimental data sets of Finotello et al.\

\subsubsection{Hydrogen}

For the solubility of molecular hydrogen, experimental data 
from various groups are available. However, different groups
report substantially different, inconsistent results.
The group of Costa Gomes~\cite{Jacquemin:2007,Gomes:2007} has studied $\rm [C_2mim][NTf_2]$, $\rm [C_4mim][NTf_2]$ and $\rm [C_6mim][NTf_2]$
and found decreasing solubility with increasing temperature.
Dyson et\,al.~\cite{Dyson:2003} just published one value for $\rm [C_4mim][NTf_2]$. 
In stark contrast to the findings of Costa Gomes,
Finotello et\,al.~\cite{Finotello:2008}, who examined $\rm [C_2mim][NTf_2]$ and $\rm [C_6mim][NTf_2]$, found a strongly 
increasing solubility with increasing temperature. 
Kume{\l}an et\,al.~\cite{Kumelan2:2006} observed this trend as well for 
$\rm [C_6mim][NTf_2]$, albeit with
a significantly weaker temperature dependence. The experimental
data of Raeissi et\,al.~\cite{Raeissi:2011} seem to match almost exactly the data of Kume{\l}an et\,al.\
Shi et\,al.~\cite{Shi:2010} published Henry constants for hydrogen
in $\rm [C_6mim][NTf_2]$ from computer simulations. Their results
support our result of a positive
slope of the temperature dependent solubility data and 
match very well the values of Kume{\l}an et\,al.~\cite{Kumelan2:2006} 
as well as  Raeissi et\,al.~\cite{Raeissi:2011}.

\subsection{Alkane Chain-Length Dependence}

\begin{figure}
  \centering 
  \includegraphics[width=7.3cm]{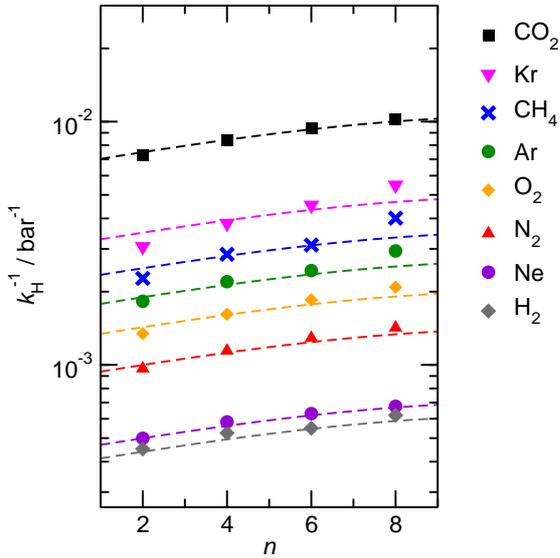}
  \caption{Comparison of inverse Henry constants of all simulated gases in [C$_n$mim][NTf$_2$] with varying chain lengths $n\!=\!2,4,6,8$ at $400\,\mbox{K}$. The dashed lines are predictions of the chain-length dependence based on the density-scaling 
  procedure described by Equation \ref{eq:scale}.}	
    \label{fig:henry_cn_lg}
\end{figure}
The computed solubilities as a function of the alkane chain length obtained from MD simulation
are given as inverse Henry coefficients, and are shown as full symbols 
in Figure \ref{fig:henry_cn_lg}.
The data indicate a rather small variation for ILs
with varying chain-length. However, there is a
significant tendency towards higher solubilities for gases in ionic liquids with longer alkane chains. When 
comparing the solubility data for $\rm [C_2mim][NTf_2]$  
and $\rm [C_8mim][NTf_2]$ for all the investigated gases, we consistently observe an increase in solublity 
of about 30\% to 40\% for the component with the C8-chain.
In Ref. \cite{Kerle:2009} we reported the observation that the solvation free energies $\mu_{\rm ex}$ of carbon
dioxide showed almost no chain-length dependence at a given temperature.
By assuming $\mu_{\rm ex}$ to be chain-length independent, 
it follows that the chain-length dependence of the solublity data at a given temperature can be solely expressed
due to density scaling according to Equation \ref{eq:solubility}, leading to
an approximate expression:
\begin{equation}
k_{\rm H}^{-1}(\rho'_{\rm IL})\!\approx\!k_{\rm H}^{-1}(\rho_{\rm IL})\times\frac{\rho_{\rm IL}}{\rho'_{\rm IL}}\;,
\label{eq:scale}
\end{equation}
where $\rho_{\rm IL}$ and $\rho'_{\rm IL}$ represent the number densities of the ion pairs in ionic liquids
with different alkane chain lengths.
Using the density data of our simulated ionic liquids 
(data is given in Ref. \cite{Kerle:2009}), we have 
fitted the number density as a function of chain length $n$ to
a second order polynomial:
\begin{equation}
\rho_{\rm IL}(n) = \rho_{\rm IL}^{(0)} + \rho_{\rm IL}^{(1)}\cdot n + \rho_{\rm IL}^{(2)}\cdot n^2
\label{eq:rhofit}
\end{equation}
with $\rho_{\rm IL}^{(0)}=2.406\,\mbox{nm}^{-3}$, $\rho_{\rm IL}^{(1)}=-0.1584\,\mbox{nm}^{-3}$, and
$\rho_{\rm IL}^{(2)}=6.39\times 10^{-3}\,\mbox{nm}^{-3}$ for $T=400\,\mbox{K}$.
As shown in Figure \ref{fig:henry_cn_lg}, the rather simple density scaling procedure describes rather accurately the chain-length dependence of the solubulity-data of
the entire set of gases, suggesting that the condition 
$\mu_{\rm ex}(n)\approx \mbox{const.}$ is mostly fulfilled for those gases.

\subsection{Temperature Dependence: ``Solvation Funnel''}

\begin{figure*}[ht!]
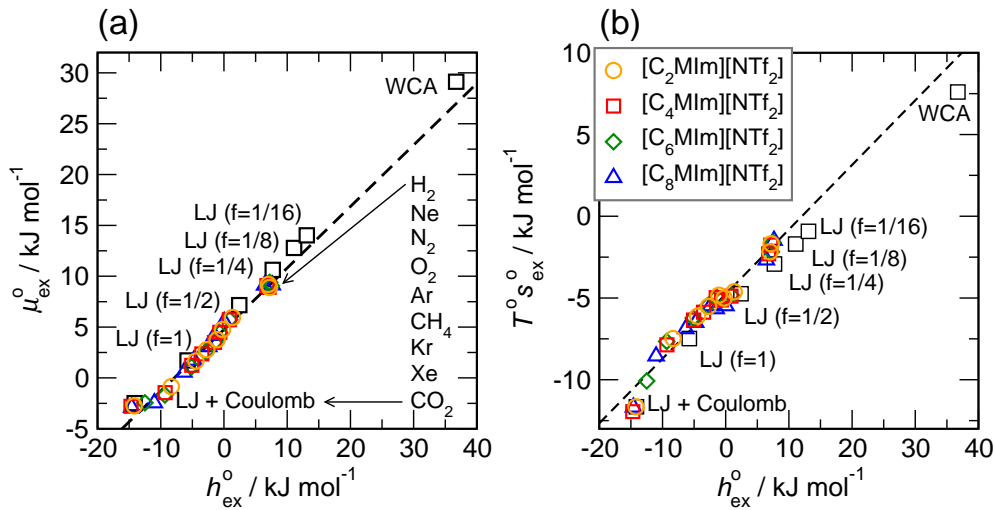

  \centering
  \footnotesize
  \centering
  \includegraphics[width=6.4cm]{FIG04a.eps}
  \includegraphics[width=6.5cm]{FIG04b.eps}

  \caption{(a) Correlation between the solvation free energy
    $\mu^\circ_{\rm ex}$ and the solvation enthalpy $h^\circ_{\rm ex}$ 
    for gases
    dissolved in [C$_n$mim][NTf$_2$] at the reference state. 
    (b)
    Correlation between the solvation entropy
    $s^\circ_{\rm ex}$ and the solvation enthalpy $h^\circ_{\rm ex}$.
    Shown are data
    for all other examined gases (colored) 
    as well as the scaled potential 
    model variants of $\mbox{CO}_2$ (black squares), taken from Ref. \cite{Kerle:2009}}
    \label{fig:simha2}
\end{figure*}
\begin{figure}
  \centering
  \includegraphics[width=7.2cm]{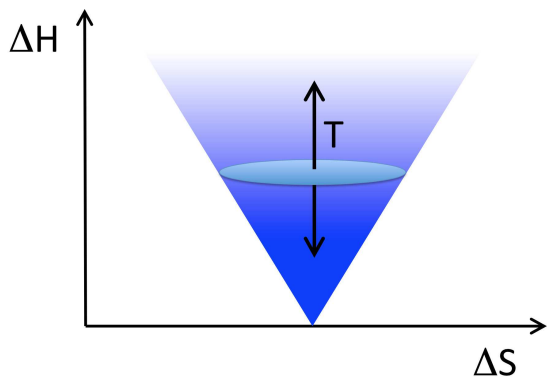}
  \caption{Schemetical representation of a hypothetical funnel-like free energy landscape of
    a light gas molecule dissolved in an ionic liquid. The gas molecule explores
    few low energy states and many high energy states,
    leading to a positive correlation between the solvation free energy and the solvation
    enthalpy.}
    \label{fig:landscape}
\end{figure}

From the wealth of experimental and
theoretical solubility data presented in Figure \ref{fig:solubility}
(including our data)
we conclude that there apparently exists a systematic relation 
between the temperature dependent slope
of the solubilities of different gases and their
interaction-strength with the solvent:
Rather ``strongly'' interacting species, such as $\mbox{CO}_2$, show
apparently a strong ``anomalous'' temperature dependence of the solubility, whereas
weakly interacting species such as oxygen and argon show a significantly weaker
temperature behavior. Finally, for the case of molecular
hydrogen $\mbox{H}_2$, we observe a change in sign of the slope, showing
a strongly increasing solubility with increasing temperature. The latter
 finding is apparently supported by the majority of experimental data sets.
However, for two cases, namely nitrogen and hydrogen, there are
substantially conflicting results from different sources, each suggesting a
very different kind
of temperature behavior.  How could this be resolved? 
We think that the apparent systematic trend is deeply rooted in
the free energy landscape explored by a solvated gas molecule,
and that the observed trend
can be explained by purely thermodynamic means.

By fitting the
temperature dependent solvation free energies $\mu_{\rm ex}(T)$
to the second order expansion around
a thermodynamic reference state 
($T^\circ\!=\!298\,\mbox{K}$ and $P^\circ\!=\!1\,\mbox{bar}$)
given by Eq. \ref{eq:cpfit}, we obtain
standard solvation free energies $\mu^\circ_{\rm ex}$, 
entropies $s^\circ_{\rm ex}$, enthalpies $h^\circ_{\rm ex}$, as
well as the solvation heat capacities $c^\circ_{P,\rm ex}$.
Data computed for all studied gases and ionic liquids are collected
in Table \ref{tab:chempot_fit}. In addition, we also consider data for
a modfied $\mbox{CO}_2$ molecule, where we have systematically weakend the
interaction with the solvent by switching off the Coulomb interaction,
and by scaling the Lennard-Jones interaction using a factor $f$ with 
$\varepsilon_{ij}=(f\varepsilon_{ii}\varepsilon_{jj})^{1/2}$. Finally, this 
prototypical molecule
is transformed into a purely repulsive component by modelling the 
solute-solvent interaction solely via Weeks-Chandler-Andersen-type (WCA) interactions
according to $V_{ij}(r)=V_{ij,\rm LJ}(r)+\varepsilon_{ij}$ for $r\leq r_{\rm LJ, min}$,
and $V_{ij}(r)=0$ otherwise. 
Following the procedure suggested by Simha et al.\ for the dissolution
of gases in polymers \cite{Simha:1969,Xie:1997},
Figure \ref{fig:simha2}a shows a plot of  the standard solvation free energy 
$\mu_{\rm ex}^\circ$ vs. the solvation enthalpy $h_{\rm ex}^\circ$
for all studied gases in all solvents.
It is evident, that both properties are linearly related 
for the entire set of solvation data. This linear relationship  can 
be utilized to predict the temperature dependent solvation data.
Furthermore, it can
provide us with a
quantitative representation of the notion that the absolute solubility of
a certain compound and its temperature behavior are somehow related.
The thermodynamic definition of the free energy implies that
the solvation entropy and enthalpy have to be linearly related as well,
as it is demonstrated in Figure \ref{fig:simha2}b.
To quantify the relations shown in Figure \ref{fig:simha2}
we use the following relation:
\begin{eqnarray}
 h^\circ_{\rm ex} &=& a\cdot \mu^\circ_{\rm ex} +b \;,
 \label{eq:linear}
\end{eqnarray} 
with $a\!=\!1.653$ and $b\!=\!-7.87\,\mbox{kJ}\,\mbox{mol}^{-1}$,
representing the parameters used for plotting the
dashed lines shown in Figure \ref{fig:simha2}a and \ref{fig:simha2}b.
From equation \ref{eq:linear} follows the
relation between $s^\circ_{\rm ex}$ and $h^\circ_{\rm ex}$ as
\begin{eqnarray}
T^\circ s_{\rm ex} = h^\circ_{\rm ex}\cdot \frac{a-1}{a} + \frac{b}{a}\;.
\end{eqnarray}
The excellent correlation between $s^\circ_{\rm ex}$ and $h^\circ_{\rm ex}$
over a rather wide range of interaction strengths is apparently
intimately related to the process of solvation of small gas molecules.
In particular, how a gas molecule and its solvation shell
explore the configurational space
of the solvated state.
To rationalize this behavior,
Figure \ref{fig:landscape} provides a
graphical representation of a hypothesized funnel-like free energy landscape of
a solvated light gas molecule.
Consequently, we call this
concept ``solvation funnel'', not unlike the ``folding funnel'' used to describe
the free energy landscape and the thermodynamic behavior of
organised polymers such as proteins \cite{Onuchic:1997}.  
The  solvated gas molecule is thought to explore
few low energy states and many high energy states, leading
to a funnel-like free energy landscape.
The consequence is a positive correlation between the solvation entropy and the solvation
enthalpy. By scaling up the interaction of the gas with the solvent, low energy states
are more strongly weighted, hence the funnel is deepened. Changing the temperature,
basically changes the population of states in the funnel-like landscape.
We would like to point out that by restricting our
study to ``small gas molecules'', the variation in size of the molecules has
apparently no big effect, and is being accounted for effectively. For larger solutes
this might not necessarily be the case.

\begin{figure}
  \centering
  \footnotesize
  \centering
  \includegraphics[width=6.4cm]{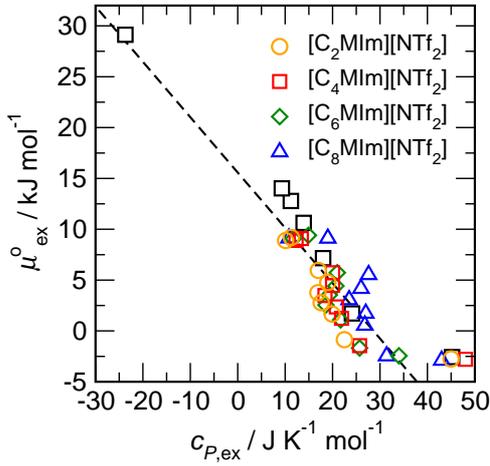}

  \caption{Correlation between the solvation  free energy
    $\mu^\circ_{\rm ex}$ and the solvation heat capacity $c_{P,\rm ex}$
    dissolved in [C$_n$mim][NTf$_2$]
    obtained for the different potential model 
    variants of $\mbox{CO}_2$ (black) \cite{Kerle:2009}
    and for all other examined gases (colored) at standard conditions.}	
    \label{fig:simhax}
\end{figure}

\begin{figure}
  \centering 
  \includegraphics[width=7.4cm]{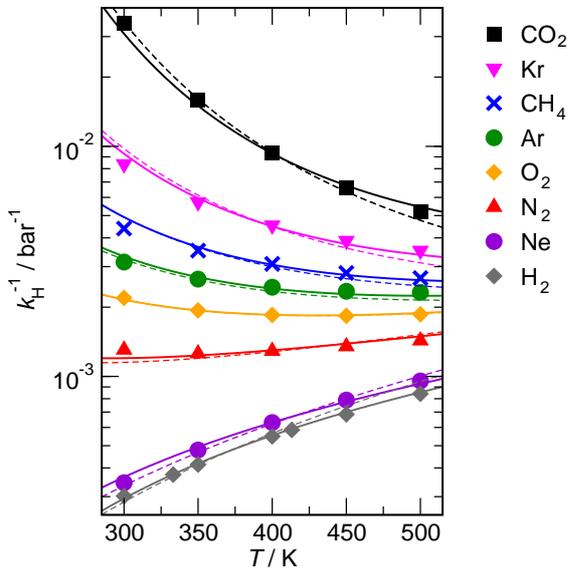}
  \caption{The symbols denote the 
   temperature dependence of the inverse Henry constants for all 
   simulated gases dissolved in $\rm [C_6mim][NTf_2]$. The lines
   indicate the predictions of the ``solvation funnel'' model. Thin dashed lines:
   $c_{P,\rm ex}=20\, \mbox{J}\,\mbox{K}^{-1}\mbox{mol}^{-1}$. Heavy 
   solid lines: $c_{P,\rm ex}$ according to Eq. \ref{eq:cpex}.}	
    \label{fig:henry_C6_vgl}
\end{figure}

The relation between $s^\circ_{\rm ex}$ and $\mu^\circ_{\rm ex}$
as outlined above,
essentially determines the coupling between the absolute solubility and
its temperature dependence. However, the second order expansion of $\mu_{ex}(T)$
 given in Eq. \ref{eq:cpfit} requires also the knowledge of the heat capacity of
solvation $c_{P,\rm ex}$, to fully determine the temperature dependence of
all our solubility data. Fortunately, the variation of
the computed $c_{P,\rm ex}$-values, 
given in Table \ref{tab:chempot_fit}, and shown in Figure \ref{fig:simhax},
has no big effect. If we neglect the heat capacity contribution completely by setting
$c_{P,\rm ex}=0$, we arrive at a description, which is
qualitatively correct
for all investigated gases (not shown). This procedure, however, leads to significant
deviations of the predicted data from our simulation data
 for temperatures above $400\,\mbox{K}$. A much better description is
 achieved by using a common value
 of $c_{P,\rm ex}=20\, \mbox{J}\,\mbox{K}^{-1}\mbox{mol}^{-1}$ for all gases instead,
 as it is indicated by the predictions represented by
 thin dashed lines in Figure \ref{fig:henry_C6_vgl}.
However, to improve things further, we make use of a negative correlation between
$\mu^\circ_{\rm ex}$  and $c_{P,\rm ex}$, suggested by Figure \ref{fig:simhax}.
To complete our ``solvation-funnel''-model 
we make use of the linear relationship
between $\mu^\circ_{\rm ex}$  and $c_{P,\rm ex}$:
\begin{eqnarray}
 c_{P,\rm ex}& = &c\cdot\mu^\circ_{\rm ex} + d
 \label{eq:cpex}
\end{eqnarray}
with $c\!=\!-1.837\cdot10^{-3}\,\mbox{K}^{-1}$ and 
      $d\!=\!28.621\,\mbox{J}\,\mbox{mol}^{-1}\,\mbox{K}^{-1}$.
With just four parameters, it is now possible to quantitatively
predict the temperature dependence of the solubility 
of all studied gases in all four solvents. The only requirement is
the knowledge of 
the solubility of a gas at
 the reference temperature $T^\circ$. 
 These
 ``solvation funnel'' model
 predictions with variable $c_{P,\rm ex}$
 are represented by thick solid lines in
 Figure \ref{fig:henry_C6_vgl}
 for all studied gases.
\begin{figure}
  \centering
  \footnotesize
  \centering
  \includegraphics[angle=0,width=8.6cm]{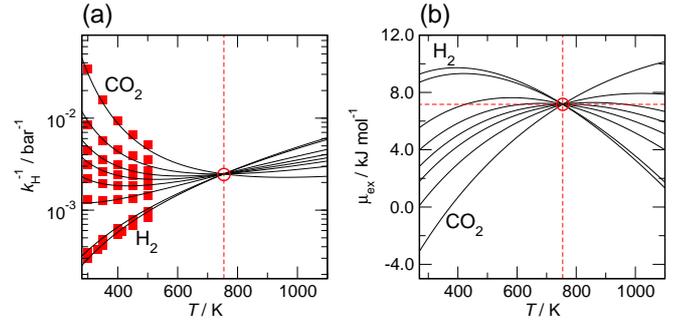}
  \caption{``Solvation funnnel'' model predictions
            for all gases in $\rm [C_6mim][NTf_2]$ with a constant 
            $c_{P,\rm ex}=20\, \mbox{J}\,\mbox{K}^{-1}\mbox{mol}^{-1}$.
            Indicated is the intersection temperature 
            $T^*=754.4\,\mbox{K}$.
             (a) Predicted solubilities. Red squares indicated experimental
             solubilities.
             (b) Predicted solvation free energies
              $\mu_{\rm ex}(T)$.
                }	
    \label{fig:commontemp}
\end{figure}
In addition to the temperature dependent 
experimental data, 
we have included
predictions according to our ``solvation funnel'' model 
also in Figure \ref{fig:solubility}. 
The thick dashed lines in 
Figure \ref{fig:solubility} represent data chosen to match
our simulation data. The thin dashed lines are used to 
illustrate the
temperature evolution of the model predictions in the vicinity of
one particular solute. Here reference solubilities
were chosen to be half and twice the size of the reference-solubility
used for matching our MD simulation data. It is quite evident that
particularly for $\rm N_2$, and $\rm H_2$ several experimental
data-sets are
incompatible with our model predictions. However, 
the red dashed lines
shown in Figure \ref{fig:solubility}
demonstrate that the experimental data obtained by the Maurer group in Kaiserslautern
(the data of Kume{\l}an et al.\ are represented by
red symbols in 
Figure \ref{fig:solubility}a,\ref{fig:solubility}b,\ref{fig:solubility}d, and  
\ref{fig:solubility}f)
for a variety of solutes are consistently in very good agreement with 
the predictions of the ``solvation funnel'' model. This includes 
even the controversial case of molecular hydrogen.

Finally, we would put another argument forward, that 
a positive slope for the temperature dependence of the
hydrogen-solubility data is a very likely scenario.
Following the arguments of Hayduk and Laudie~\cite{Hayduk:1973}, which have
been reviewed and extended by Beutier and Renon~\cite{Beutier:1978}, 
all Henry constants and hence all solubilities obtained for a certain solvent
should meet at the critical 
point of that solvent, in our case the ionic liquids. 
The solubility data shown in Figure \ref{fig:henry_C6_vgl} clearly indicate
a convergent behavior  with increasing temperature. 
By extrapolating 
the ``solvation funnel'' model to very high temperatures, we
find that this convergence even continues. 
In addition, Figure \ref{fig:commontemp}
demonstrates that by assuming $c_{P,\rm ex}$ to be constant,
the ``solvation funnel'' model even predicts 
a single common intersection temperature $T^*$ for 
the solubility all gases.
Equations \ref{eq:cpfit} and \ref{eq:linear} imply
that the common temperature $T^*$ is defined by
\begin{equation}
T^* = T^\circ \cdot \frac{a}{a-1}
\label{eq:tcommon}
\end{equation}
with the corresponding common solvation free energy of
\begin{equation}
\mu_{\rm ex}(T^*)
=
\frac{b}{1-a} + \frac{c_{P,\rm ex}T^\circ}{1-a}
\left[
a\cdot \ln \left( \frac{a}{a-1}\right) -1
\right]\;.
\end{equation}
It is remarkable that Equation \ref{eq:tcommon} just 
requires the knowledge of the parameter $a$,
defined in  Equation \ref{eq:linear}, as well as
the chosen reference temperature $T^\circ$.
By using $a=1.653$ and $T^\circ=298\,\mbox{K}$, we compute
a common temperature of  $T^*=754.4\,\mbox{K}$ for 
all gases used in this study. The solubility curves intersect at a value of
$\mu_{\rm ex}(T^*)=7.166 \,\mbox{kJ}\,\mbox{mol}^{-1}$
for the case of $c_{P, \rm ex}\!=\!20\,\mbox{J}\,\mbox{K}^{-1}\mbox{mol}^{-1}$
(indicated in Figure \ref{fig:commontemp}). 
However, by allowing $c_{P,\rm ex}$ to vary slightly for each gas,
this constraint of a common temperature is violated.
This does not necessarily mean that our model
is inadequate, but rather that the temperature range of our study is too
far away from the critical temperature.
It is not unlikely that, by approaching the critical
point, the solvation heat capacities $c_{P,\rm ex}$ will eventually converge,
while the solvated gases maintain their solvation enthalpy-entropy correlation.
This would again restore the common temperature feature.
Although the predicted solubility-data with variable $c_{P,\rm ex}$
do not meet exactly at a particular temperature, they show
a region of nearest approximation around $1100\,\mbox{K}$,
 which is far above the decomposition point of any of
 the ionic liquids. 
 We would like to point out, however, that this
 coincides nicely
 with results from Rebelo et\,al.~\cite{Rebelo:2005} and Freire et\,al.~\cite{Freire:2007}, who estimated the location of 
 the critical temperature to be around $1100\,\mbox{K}$
 using  surface tension data in combination with
 the E\"{o}tvos and Guggenheim equations. Their predicted critical temperature, 
 however, was significantly
 larger than the predictions of
 Yokozeki et\,al.~\cite{Yokozeki:2008}, who used 
 the Vetere method, as well as of Shin et\,al.~\cite{Shin:2008},
  who used the group contribution method (GCM) of Joback, and
  Valderrama et\,al.~\cite{Valderrama:2007}, who obtained
  results applying the modified Lydersen-Joback-Reid method (mLJR).

\section{Conclusion}

The systematic behavior of the gas solubility in Ionic Liquids is
studied and described with the help of extensive molecular dynamics
simulations. 
The solubility of hydrogen, oxygen, nitrogen, methane, krypton, 
argon, neon and carbon dioxide in the Ionic Liquids of type
1-n-alkyl-3-methylimidazolium bis(trifluoromethylsulfonyl)imide
([C$_n$mim][NTf$_2$]) with varying chain lengths $n\!=\!2,4,6,8$ 
are computed  for a large temperature range from 
$300\,\mbox{K}$ up to $500\,\mbox{K}$ at $1\,\mbox{bar}$.
By applying Widom's particle insertion technique,
as well as Bennett's overlapping distribution method,
we are able to determine
solvation free energies for those selected light gases in
imidazolium based Ionic Liquids with great
statistical accuracy. 
A detailed comparison
of the computed solubility data with available 
experimental and theoretical data is provided.

We observe, that the chain-length dependence of the  computed solubility 
in various solvents can be mostly attributed
to the change of the number-density of ion-pairs in the solvent,
as the computed solvation free energies show almost no
chain-length dependence.

The  data
obtained from our MD simulations clearly
show that the magnitude of the solvation free energy 
at a defined reference temperature and
its temperature-derivatives are intimately related with respect to
one-another. 
This is a consequence of a  solvation entropy-enthalpy
compensation effect, which seems to be a defining feature 
of the solvation  of small molecules in 
the investigated Ionic Liquids.
We rationalize  this feature as the consequence of
a hypothesized funnel-like free-energy landscape
explored by the solvated gas molecule.
This effect is leading to simple analytical relations
quantitatively
describing the temperature dependent solubility of gases
solely depending on the absolute solubility value at
a defined reference temperature which we call 
``solvation funnel'' model.

The ``solvation funnel'' model  is also predicting
that the solubility data all meet at a  single temperature,
which is in line with the observation made for 
various fluids that the solubilities for gases meet
at the critical temperature. However, this feature of
a common temperature exists only, if the
model is used with a unique heat capacity of solvation
valid for all gases. Since the computed
heat capacities of solvation do not vary strongly for
different gases,
a common value of 
$c_{P, \rm ex}\!=\!20\,\mbox{J}\,\mbox{K}^{-1}\mbox{mol}^{-1}$
is a reasonable approximation for all 
investigated solutes.

We would like to point out that the ``solvation funnel'' model
is particularly helpful for assessing the solvation
behavior of very light gases, such as hydrogen, where conflicting
experimental data have been reported.

\section{Acknowledgements}

This work has been supported by the German 
Science Foundation (DFG) priority
program SPP 1570 with additional support from SFB 652.


\begin{thebibliography}{100}
\expandafter\ifx\csname urlstyle\endcsname\relax 
  \providecommand{\doi}[1]{DOI \discretionary{}{}{}#1}\else 
  \providecommand{\doi}{DOI \discretionary{}{}{}\begingroup 
  \urlstyle{rm}\Url}\fi 
\providecommand{\bibAnnoteFile}[1]{%
  \IfFileExists{#1}{\begin{quotation}\noindent\textsc{Key:} #1\\
  \textsc{Annotation:}\ \input{#1}\end{quotation}}{}}
\providecommand{\bibAnnote}[2]{%
  \begin{quotation}\noindent\textsc{Key:} #1\\
  \textsc{Annotation:}\ #2\end{quotation}}

\bibitem{WasserscheidWelton}
P.~Wasserscheid, T.~Welton, \emph{Ionic Liquids in Synthesis}, 2nd Aufl.,
  VCH-Wiley, Weinheim, \textbf{2007}. 
\input{WasserscheidWelton}

\bibitem{Angell:2012}
C.~A. Angell, Y.~Ansari, Z.~Zhao, \emph{Faraday Discuss.} \textbf{2012},
  \emph{154}, 9--27. 
\input{Angell:2012}

\bibitem{Endres:2006}
F.~Endres, S.~Z. {El Abedin}, \emph{Phys. Chem. Chem. Phys.} \textbf{2006},
  \emph{8}, 2101--2116. 
\input{Endres:2006}

\bibitem{Rogers:2003}
R.~D. Rogers, K.~R. Seddon, \emph{Science} \textbf{2003}, \emph{302}, 792--793. 
\input{Rogers:2003}

\bibitem{Smiglak:2007}
M.~Smiglak, A.~Metlen, R.~D. Rogers, \emph{Acc. Chem. Res.} \textbf{2007},
  \emph{40}, 1182--1192. 
\input{Smiglak:2007}

\bibitem{Plechkova:2008}
N.~V. Plechkova, K.~R. Seddon, \emph{Chem. Rev.} \textbf{2008}, \emph{37},
  123--150. 
\input{Plechkova:2008}

\bibitem{Blanchard:1999}
L.~A. Blanchard, D.~Hancu, E.~J. Beckman, J.~F. Brennecke, \emph{Nature}
  \textbf{1999}, \emph{399}, 28--29. 
\input{Blanchard:1999}

\bibitem{Tempel:2008}
D.~J. Tempel, P.~B. Henderson, J.~R. Brzozowski, R.~M. Pearlstein, H.~Cheng,
  \emph{J. Am. Chem. Soc.} \textbf{2008}, \emph{130}, 400--401. 
\input{Tempel:2008}

\bibitem{Huang:2009}
J.~Huang, T.~R\"uther, \emph{Aust. J. Chem.} \textbf{2009}, \emph{62},
  298--308. 
\input{Huang:2009}

\bibitem{Ramdin:2012}
M.~Ramdin, T.~W. {de Loos}, T.~J.~H. Vlugt, \emph{Ind. Eng. Chem. Res.}
  \textbf{2012}, \emph{51}, 8149--8177. 
\input{Ramdin:2012}

\bibitem{Bara:2010}
J.~E. Bara, D.~E. Camper, D.~L. Gin, R.~D. Noble, \emph{Acc. Chem. Res.}
  \textbf{2010}, \emph{43}, 152--159. 
\input{Bara:2010}

\bibitem{Raeissi:2009}
S.~Raeissi, C.~J. Peters, \emph{Green Chem.} \textbf{2009}, \emph{11},
  185--192. 
\input{Raeissi:2009}

\bibitem{Myers:2008}
C.~Myers, H.~Pennline, D.~Luebke, J.~Ilconich, J.~K. Dixon, E.~J. Maginn, J.~F. 
  Brennecke, \emph{J. Membrane Science} \textbf{2008}, \emph{322}, 28--31. 
\input{Myers:2008}

\bibitem{Ilconich:2007}
J.~Ilconich, C.~Myers, H.~Pennline, D.~Luebke, \emph{J. Membrane Science}
  \textbf{2007}, \emph{298}, 41--47. 
\input{Ilconich:2007}

\bibitem{Lei:2013}
Z.~Lei, C.~Dai, B.~Chen, \emph{Chem. Rev.} \textbf{2013}, \emph{0}, null,
  \doi{10.1021/cr300497a}. 
\input{Lei:2013}

\bibitem{Finotello:2008}
A.~Finotello, J.~E. Bara, D.~Camper, R.~D. Noble, \emph{Ind. Eng. Chem. Res.}
  \textbf{2008}, \emph{47}, 3453--3459. 
\input{Finotello:2008}

\bibitem{Jacquemin:2007}
J.~Jacquemin, P.~Husson, V.~Majer, M.~F. {Costa Gomes}, \emph{J. Solution 
  Chem.} \textbf{2007}, \emph{36}, 967--979. 
\input{Jacquemin:2007}

\bibitem{Gomes:2007}
M.~F.~C. Gomes, \emph{J. Chem. Eng. Data} \textbf{2007}, \emph{52}, 472--475. 
\input{Gomes:2007}

\bibitem{Kumelan1:2006}
J.~Kume{\l}an, A.~P.~S. Kamps, D.~Tuma, G.~Maurer, \emph{J. Chem. 
  Thermodynamics} \textbf{2006}, \emph{38}, 1396--1401. 
\input{Kumelan1:2006}

\bibitem{Dyson:2003}
P.~J. Dyson, G.~Laurenczy, C.~A. Ohlin, J.~Vallance, T.~Welton, \emph{Chem. 
  Commun} \textbf{2003}, \emph{.}, 2418--2419. 
\input{Dyson:2003}

\bibitem{Jacquemin:2006}
J.~Jacquemin, M.~F. {Costa Gomes}, P.~Husson, V.~Majer, \emph{J. Chem. 
  Thermodynamics} \textbf{2006}, \emph{38}, 490--502. 
\input{Jacquemin:2006}

\bibitem{Jacquemin:2008}
J.~Jacquemin, P.~Husson, V.~Majer, A.~A.~H. Padua, M.~F. {Costa Gomes},
  \emph{Green Chem.} \textbf{2008}, \emph{10}, 944--950. 
\input{Jacquemin:2008}

\bibitem{Jacquemin2:2006}
J.~Jacquemin, P.~Husson, V.~Majer, M.~F. {Costa Gomes}, \emph{Fluid Phase 
  Equilibria} \textbf{2006}, \emph{1}, 87--95. 
\input{Jacquemin2:2006}

\bibitem{Kumelan3:2006}
J.~Kume{\l}an, A.~P.~S. Kamps, D.~Tuma, G.~Maurer, \emph{J. Chem. Eng. Data}
  \textbf{2006}, \emph{51}, 11--14. 
\input{Kumelan3:2006}

\bibitem{Kumelan:2010}
J.~Kume{\l}an, D.~Tuma, A.~P.~S. Kamps, G.~Maurer, \emph{J. Chem. Eng. Data.}
  \textbf{2010}, \emph{55}, 165--172. 
\input{Kumelan:2010}

\bibitem{Raeissi:2011}
S.~Raeissi, L.~J. Florusse, C.~J. Peters, \emph{J. Chem. Eng. Data}
  \textbf{2011}, \emph{56}, 1105--1107. 
\input{Raeissi:2011}

\bibitem{Shimoyama:2010}
Y.~Shimoyama, A.~Ito, \emph{Fluid Phase Equilibria} \textbf{2010}, \emph{297},
  178--182. 
\input{Shimoyama:2010}

\bibitem{Manan:2009}
N.~A. Manan, C.~Hardacre, J.~Jacquemin, D.~W. Rooney, T.~G.~A. Youngs, \emph{J. 
  Chem. Eng. Data} \textbf{2009}, \emph{54}, 2005--2022. 
\input{Manan:2009}

\bibitem{Palomar:2011}
J.~Palomar, M.~Gonzalez-Miquel, A.~Polo, F.~Rodriguez, \emph{Ind. Eng. Chem. 
  Res.} \textbf{2011}, \emph{50}, 3452--3463. 
\input{Palomar:2011}

\bibitem{Sumon:2011}
K.~Z. Sumon, A.~Henni, \emph{Fluid Phase Equilibria} \textbf{2011}, \emph{310},
  39--55. 
\input{Sumon:2011}

\bibitem{Abildskov:2009}
J.~Abildskov, M.~D. Ellegaard, J.~P. O'Connell, \emph{Fluid Phase Equilibria}
  \textbf{2009}, \emph{286}, 95--106. 
\input{Abildskov:2009}

\bibitem{Andreu:2008}
J.~S. Andreu, L.~F. Vega, \emph{J. Phys. Chem. B} \textbf{2008}, \emph{112},
  15398--15406. 
\input{Andreu:2008}

\bibitem{Zhang:2008}
X.~Zhang, Z.~Liu, W.~Wang, \emph{AlChE} \textbf{2008}, \emph{54}, 2717--2728. 
\input{Zhang:2008}

\bibitem{Andreu:2007}
J.~S. Andreu, L.~F. Vega, \emph{J. Phys. Chem. C} \textbf{2007}, \emph{111},
  16028--16034. 
\input{Andreu:2007}

\bibitem{Kim:2007}
Y.~S. Kim, J.~H. Jang, B.~D. Lim, J.~W. Kang, C.~S. Lee, \emph{Fluid Phase 
  Equilibria} \textbf{2007}, \emph{256}, 70--74. 
\input{Kim:2007}

\bibitem{Oliferenko:2011}
A.~A. Oliferenko, P.~Oliferenko, K.~R. Seddon, J.~S. Torrecilla, \emph{Phys. 
  Chem. Chem. Phys.} \textbf{2011}, \emph{13}, 17262--17272. 
\input{Oliferenko:2011}

\bibitem{Maginn:2009}
E.~J. Maginn, \emph{J. Phys.: Condens. Matter} \textbf{2009}, \emph{21},
  373101--37318. 
\input{Maginn:2009}

\bibitem{Shi1:2008}
W.~Shi, E.~J. Maginn, \emph{J. Phys. Chem. B} \textbf{2008}, \emph{112},
  2045--2055. 
\input{Shi1:2008}

\bibitem{Shi2:2008}
W.~Shi, E.~J. Maginn, \emph{J. Phys. Chem. B} \textbf{2008}, \emph{112},
  16710--16720. 
\input{Shi2:2008}

\bibitem{Koeddermann:2008a}
T.~K\"oddermann, D.~Paschek, R.~Ludwig, \emph{ChemPhysChem} \textbf{2008},
  \emph{9}, 549--555. 
\input{Koeddermann:2008a}

\bibitem{Lynden-Bell:2002}
R.~M. Lynden-Bell, N.~A. Atamas, A.~Vasilyuk, C.~G. Hanke, \emph{Mol. Phys.}
  \textbf{2002}, \emph{100}, 3225--3229. 
\input{Lynden-Bell:2002}

\bibitem{Hanke:2003}
C.~G. Hanke, A.~Johansson, J.~B. Harper, R.~M. Lynden-Bell, \emph{Chem. Phys. 
  Lett.} \textbf{2003}, \emph{374}, 85--89. 
\input{Hanke:2003}

\bibitem{Deschamps:2004}
J.~Deschamps, M.~F. {Costa Gomes}, A.~A.~H. Padua, \emph{ChemPhysChem}
  \textbf{2004}, \emph{5}, 1049--1052. 
\input{Deschamps:2004}

\bibitem{Shah:2004}
J.~K. Shah, E.~J. Maginn, \emph{Fluid Phase Equilibria} \textbf{2004},
  \emph{195}, 222--223. 
\input{Shah:2004}

\bibitem{Shah:2005}
J.~K. Shah, E.~J. Maginn, \emph{J. Phys. Chem. B} \textbf{2005}, \emph{109},
  10395--10405. 
\input{Shah:2005}

\bibitem{Cadena:2004}
C.~Cadena, J.~L. Anthony, J.~K. Shah, T.~I. Morrow, J.~F. Brennecke, E.~J. 
  Maginn, \emph{J. Am. Chem. Soc.} \textbf{2004}, \emph{126}, 5300--5308. 
\input{Cadena:2004}

\bibitem{Kumelan:2005}
J.~Kume{\l}an, A.~P.~S. Kamps, I.~Urukova, D.~Tuma, G.~Maurer, \emph{J. Chem. 
  Thermodynamics} \textbf{2005}, \emph{37}, 595--602. 
\input{Kumelan:2005}

\bibitem{Urukova:2005}
I.~Urukova, J.~Vorholz, G.~Maurer, \emph{J. Phys. Chem. B} \textbf{2005},
  \emph{109}, 12154--12159. 
\input{Urukova:2005}

\bibitem{Kumelan:2007}
J.~Kume{\l}an, A.~P.~S. Kamps, D.~Tuma, G.~Maurer, \emph{Ind. Eng. Chem. Res.}
  \textbf{2007}, \emph{46}, 8236--8240. 
\input{Kumelan:2007}

\bibitem{Deschamps:2005}
J.~Deschamps, , A.~A.~H. P\'adua, \emph{ACS Symp. Series} \textbf{2005},
  \emph{901}, 150--158. 
\input{Deschamps:2005}

\bibitem{Wu:2005}
X.~P. Wu, Z.~P. Liu, W.~C. Wang, \emph{Wuli Huaxue Xuebao} \textbf{2005},
  \emph{21}, 1138--1142. 
\input{Wu:2005}

\bibitem{Shi:2010}
W.~Shi, D.~C. Sorescu, D.~R. Luebke, M.~J. Keller, S.~Wickramanayake, \emph{J. 
  Phys. Chem. B} \textbf{2010}, \emph{114}, 6531--6541. 
\input{Shi:2010}

\bibitem{Ghobadi:2011}
A.~F. Ghobadi, V.~Taghikhani, J.~R. Elliott, \emph{J. Phys. Chem. B}
  \textbf{2011}, \emph{115}, 13599--13607. 
\input{Ghobadi:2011}

\bibitem{Dommert:2012}
F.~Dommert, K.~Wendler, R.~Berger, L.~{Delle Site}, C.~Holm,
  \emph{ChemPhysChem} \textbf{2012}, \emph{13}, 1625--1637. 
\input{Dommert:2012}

\bibitem{Lopes:2004}
J.~N. {Canongia Lopes}, J.~Deschamps, A.~A.~H. Padua, \emph{J. Phys. Chem. B}
  \textbf{2004}, \emph{108}, 2038--2047. 
\input{Lopes:2004}

\bibitem{Koeddermann:2007}
T.~K\"oddermann, D.~Paschek, R.~Ludwig, \emph{ChemPhysChem} \textbf{2007},
  \emph{8}, 2464--2470. 
\input{Koeddermann:2007}

\bibitem{Paschek:2008}
D.~Paschek, T.~K\"oddermann, R.~Ludwig, \emph{Phys. Rev. Lett.} \textbf{2008},
  \emph{100}, 115901. 
\input{Paschek:2008}

\bibitem{Kerle:2009}
D.~Kerl\'e, R.~Ludwig, A.~Geiger, D.~Paschek, \emph{J. Phys. Chem. B}
  \textbf{2009}, \emph{113}, 12727--12735. 
\input{Kerle:2009}

\bibitem{PrattRev:2002}
L.~R. Pratt, \emph{Annu. Rev. Phys. Chem.} \textbf{2003}, \emph{53}, 409--436. 
\input{PrattRev:2002}

\bibitem{Southall:2002}
N.~T. Southall, K.~A. Dill, A.~D.~J. Haymet, \emph{J. Phys. Chem. B}
  \textbf{2002}, \emph{106}, 521--533. 
\input{Southall:2002}

\bibitem{Widom:2003}
B.~Widom, P.~Bhimalapuram, K.~Koga, \emph{Phys. Chem. Chem. Phys.}
  \textbf{2003}, \emph{5}, 3085--3093. 
\input{Widom:2003}

\bibitem{Chandler:2005}
D.~Chandler, \emph{Nature (London)} \textbf{2005}, \emph{437}, 640--647. 
\input{Chandler:2005}

\bibitem{Wittich:2010}
B.~Wittich, U.~K. Deiters, \emph{J. Phys. Chem. B} \textbf{2010}, \emph{114},
  3452--3463. 
\input{Wittich:2010}

\bibitem{Patkowski:2008}
K.~Patkowski, W.~Cencek, P.~Jankowski, K.~Szalewicz, J.~Mehl, G.~Garberoglio,
  A.~H. Harvey, \emph{J. Chem. Phys.} \textbf{2008}, \emph{129},
  094304--094323. 
\input{Patkowski:2008}

\bibitem{Guillot:93}
B.~Guillot, Y.~Guissani, \emph{J. Chem. Phys.} \textbf{1993}, \emph{99},
  8075--8094. 
\input{Guillot:93}

\bibitem{Potoff:2001}
J.~Potoff, J.~I. Siepmann, \emph{AIChE Journal} \textbf{2001}, \emph{47},
  1676--1682. 
\input{Potoff:2001}

\bibitem{Hansen:2007}
N.~Hansen, F.~A. B.Agbor, F.~J. Keil, \emph{Fluid Phase Equilibria}
  \textbf{2007}, \emph{259}, 180--188. 
\input{Hansen:2007}

\bibitem{Harris:1995}
J.~G. Harris, K.~H. Yung, \emph{J. Phys. Chem.} \textbf{1995}, \emph{99},
  12021--12024. 
\input{Harris:1995}

\bibitem{gmxpaper}
E.~Lindahl, B.~Hess, D.~{van der Spoel}, \emph{J. Mol. Model.} \textbf{2001},
  \emph{7}, 306--317. 
\input{gmxpaper}

\bibitem{MOSCITO}
D.~Paschek, {MOSCITO 4}: {MD} simulation package, \textbf{2008},
  http://ganter.chemie.uni-dortmund.de/MOSCITO. 
\input{MOSCITO}

\bibitem{Nose:1984}
S.~Nos\'e, \emph{Mol. Phys.} \textbf{1984}, \emph{52}, 255--268. 
\input{Nose:1984}

\bibitem{Hoover:1985}
W.~G. Hoover, \emph{Phys. Rev. A} \textbf{1985}, \emph{31}, 1695--1697. 
\input{Hoover:1985}

\bibitem{Parrinello:1981}
M.~Parrinello, A.~Rahman, \emph{J. Appl. Phys.} \textbf{1981}, \emph{52},
  7182--7180. 
\input{Parrinello:1981}

\bibitem{Nose:1983}
S.~Nos\'e, M.~L. Klein, \emph{Mol. Phys.} \textbf{1983}, \emph{50}, 1055--1076. 
\input{Nose:1983}

\bibitem{Essmann:1995}
U.~Essmann, L.~Perera, M.~L. Berkowitz, T.~A. Darden, H.~Lee, L.~G. Pedersen,
  \emph{J. Chem. Phys.} \textbf{1995}, \emph{103}, 8577--8593. 
\input{Essmann:1995}

\bibitem{Ryckaert:1977}
J.~P. Ryckaert, G.~Ciccotti, H.~J.~C. Berendsen, \emph{J. Comp. Phys.}
  \textbf{1977}, \emph{23}, 327--341. 
\input{Ryckaert:1977}

\bibitem{Kennan:90}
R.~P. Kennan, G.~L. Pollack, \emph{J. Chem. Phys.} \textbf{1990}, \emph{93},
  2724--2735. 
\input{Kennan:90}

\bibitem{Widom:63}
B.~Widom, \emph{J. Chem. Phys.} \textbf{1963}, \emph{39}, 2808--2812. 
\input{Widom:63}

\bibitem{Beckbook}
T.~L. Beck, M.~E. Paulaitis, L.~R. Pratt, \emph{The Potential Distribution 
  Theorem and Models of Molecular Solutions}, Cambridge University Press,
  Cambrigde, UK, \textbf{2006}. 
\input{Beckbook}

\bibitem{Bennett:1976}
C.~H. Bennett, \emph{J. Comp. Phys.} \textbf{1976}, \emph{22}, 245--268. 
\input{Bennett:1976}

\bibitem{Shing:1983}
K.~S. Shing, K.~E. Gubbins, \emph{Mol. Phys.} \textbf{1983}, \emph{49},
  1121--1138. 
\input{Shing:1983}

\bibitem{FrenkelSmit}
D.~Frenkel, B.~Smit, \emph{Understanding Molecular Simulation. From Algorithms 
  to Applications}, 2nd Aufl., Academic Press, San Diego, \textbf{2002}. 
\input{FrenkelSmit}

\bibitem{Paschek:2004:1}
D.~Paschek, \emph{J. Chem. Phys.} \textbf{2004}, \emph{120}, 6674--6690. 
\input{Paschek:2004:1}

\bibitem{Roberts:94}
J.~E. Roberts, J.~Schnitker, \emph{J. Chem. Phys.} \textbf{1994}, \emph{101},
  5024--5031. 
\input{Roberts:94}

\bibitem{Paschek:2004:2}
D.~Paschek, \emph{J. Chem. Phys.} \textbf{2004}, \emph{120}, 10605--10617. 
\input{Paschek:2004:2}

\bibitem{densityfit}
To interconvert solvation free energies $\mu_{\rm ex}$ and Henry constants 
  $k_{\rm H}$, we use a second order polynomial fitted to the temperature 
  dependent ion-pair density of the simulated ionic liquids: $\rho_{\rm 
  IL}(T)=\rho_{\rm IL}^{(0)}+\rho_{\rm IL}^{(1)}\cdot T+ \rho_{\rm 
  IL}^{(2)}\cdot T^2$. $\rm [C_2mim][NTf_2]$: $\rho_{\rm 
  IL}^{(0)}=2.985\,\mbox{nm}^{-3}$, $\rho_{\rm IL}^{(1)}=-2.51\times 
  10^{-3}\,\mbox{nm}^{-3}\,\mbox{K}^{-1}$, and $\rho_{\rm IL}^{(2)}=8.46\times 
  10^{-7}\,\mbox{nm}^{-3}\,\mbox{K}^{-2}$. $\rm [C_4mim][NTf_2]$: $\rho_{\rm 
  IL}^{(0)}=2.639\,\mbox{nm}^{-3}$, $\rho_{\rm IL}^{(1)}=-2.18\times 
  10^{-3}\,\mbox{nm}^{-3}\,\mbox{K}^{-1}$, and $\rho_{\rm IL}^{(2)}=6.85\times 
  10^{-7}\,\mbox{nm}^{-3}\,\mbox{K}^{-2}$. $\rm [C_6mim][NTf_2]$: $\rho_{\rm 
  IL}^{(0)}=2.368\,\mbox{nm}^{-3}$, $\rho_{\rm IL}^{(1)}=-1.94\times 
  10^{-3}\,\mbox{nm}^{-3}\,\mbox{K}^{-1}$, and $\rho_{\rm IL}^{(2)}=5.90\times 
  10^{-7}\,\mbox{nm}^{-3}\,\mbox{K}^{-2}$. $\rm [C_8mim][NTf_2]$: $\rho_{\rm 
  IL}^{(0)}=2.127\,\mbox{nm}^{-3}$, $\rho_{\rm IL}^{(1)}=-1.56\times 
  10^{-3}\,\mbox{nm}^{-3}\,\mbox{K}^{-1}$, and $\rho_{\rm IL}^{(2)}=2.92\times 
  10^{-7}\,\mbox{nm}^{-3}\,\mbox{K}^{-2}$. 
\input{densityfit}

\bibitem{IUPAC1}
K.~N. Marsh, J.~F. Brennecke, R.~D. Chirico, M.~Frenkel, A.~Heintz, J.~W. 
  Magee, C.~J. Peters, L.~P.~N. Rebelo, K.~R. Seddon, \emph{Pure Appl. Chem.}
  \textbf{2009}, \emph{81}, 781--790. 
\input{IUPAC1}

\bibitem{IUPAC2}
R.~D. Chirico, V.~Diky, J.~W. Magee, M.~Frenkel, K.~N. Marsh, \emph{Pure Appl. 
  Chem.} \textbf{2009}, \emph{81}, 791--828. 
\input{IUPAC2}

\bibitem{Kumelan:2006}
J.~Kume{\l}an, A.~P.~S. Kamps, D.~Tuma, G.~Maurer, \emph{J. Chem. 
  Thermodynamics} \textbf{2006}, \emph{38}, 1396--1401. 
\input{Kumelan:2006}

\bibitem{Moganty:2010}
S.~S. Moganty, R.~E. Baltus, \emph{Ind. Eng. Chem. Res.} \textbf{2010},
  \emph{49}, 5846--5853. 
\input{Moganty:2010}

\bibitem{Kerle:2013}
D.~Kerl\'e, R.~Ludwig, A.~Geiger, D.~Paschek, \emph{Z. Phys. Chem.}
  \textbf{2013}, \emph{227}, 167--176. 
\input{Kerle:2013}

\bibitem{Camper2:2006}
D.~Camper, J.~Bara, C.~Koval, R.~Noble, \emph{Ind. Eng. Chem. Res.}
  \textbf{2006}, \emph{45}, 6279--6283. 
\input{Camper2:2006}

\bibitem{Anderson:2007}
J.~L. Anderson, J.~K. Dixon, J.~F. Brennecke, \emph{Acc. Chem. Res.}
  \textbf{2007}, \emph{40}, 1208--1216. 
\input{Anderson:2007}

\bibitem{Blath:2011}
J.~Blath, M.~Christ, N.~Deubler, T.~Hirth, T.~Schiestel, \emph{Chem. Eng. J.}
  \textbf{2011}, \emph{172}, 167--176. 
\input{Blath:2011}

\bibitem{Afzal:2013}
W.~Afzal, X.~Liu, J.~M. Prausnitz, \emph{J. Chem. Thermodynamics}
  \textbf{2013}, \emph{63}, 88--94. 
\input{Afzal:2013}

\bibitem{Kumelan1:2009}
J.~Kume{\l}an, A.~P.~S. Kamps, D.~Tuma, G.~Maurer, \emph{J. Chem. Eng. Data}
  \textbf{2009}, \emph{54}, 966--971. 
\input{Kumelan1:2009}

\bibitem{Anthony:2005}
J.~L. Anthony, J.~L. Anderson, E.~J. Maginn, J.~F. Brennecke, \emph{J. Phys. 
  Chem. B} \textbf{2005}, \emph{109}, 6366--6374. 
\input{Anthony:2005}

\bibitem{Kumelan2:2006}
J.~Kume{\l}an, A.~P.~S. Kamps, D.~Tuma, G.~Maurer, \emph{J. Chem. Eng. Data}
  \textbf{2006}, \emph{51}, 1364--1367. 
\input{Kumelan2:2006}

\bibitem{Simha:1969}
R.~Simha, T.~Somcynsky, \emph{Macromolecules} \textbf{1969}, \emph{2},
  342--350. 
\input{Simha:1969}

\bibitem{Xie:1997}
H.~Xie, R.~Simha, \emph{Polymer Int.} \textbf{1997}, \emph{44}, 348--355. 
\input{Xie:1997}

\bibitem{Onuchic:1997}
J.~N. Onuchic, Z.~{Luthey-Schulten}, P.~G. Wolynes, \emph{Annu. Rev. Phys. 
  Chem.} \textbf{1997}, \emph{48}, 545--600. 
\input{Onuchic:1997}

\bibitem{Hayduk:1973}
W.~Hayduk, H.~Laudie, \emph{AIChE Journal} \textbf{1973}, \emph{19},
  1233--1238. 
\input{Hayduk:1973}

\bibitem{Beutier:1978}
D.~Beutier, H.~Renon, \emph{AIChE Journal} \textbf{1978}, \emph{24},
  1122--1125. 
\input{Beutier:1978}

\bibitem{Rebelo:2005}
L.~P.~N. Rebelo, J.~N. {Canongia Lopes}, J.~M. S.~S. Esperan\c{c}a, E.~Filipe,
  \emph{J. Phys. Chem B} \textbf{2005}, \emph{109}, 6040--6043. 
\input{Rebelo:2005}

\bibitem{Freire:2007}
M.~G. Freire, P.~J. Carvalho, A.~M. Fernandes, I.~M. Marrucho, A.~J. Queimada,
  J.~A.~P. Coutinho, \emph{J. Colloid Interf. Sci.} \textbf{2007}, \emph{314},
  621--630. 
\input{Freire:2007}

\bibitem{Yokozeki:2008}
A.~Yokozeki, M.~B. Shiflett, C.~P. Junk, L.~M. Grieco, T.~Foo, \emph{J. Phys. 
  Chem. B} \textbf{2008}, \emph{112}, 16654--16663. 
\input{Yokozeki:2008}

\bibitem{Shin:2008}
E.-K. Shin, B.-C. Lee, J.~S. Lim, \emph{J. Supercrit. Fluids.} \textbf{2008},
  \emph{45}, 282--292. 
\input{Shin:2008}
\bibitem{Valderrama:2007}
J.~O. Valderrama, P.~A. Robles, \emph{Ind. Eng. Chem. Res} \textbf{2007},
  \emph{46}, 1338--1344. 
\input{Valderrama:2007}

\end{thebibliography}

\end{document}